%
%
%
%
%

\input harvmac
\input epsf

\parindent=0pt
\Title{SHEP 99-09}{\vbox{\centerline{A gauge invariant exact 
renormalization group I}}}

\centerline{\bf Tim R. Morris}
\vskip .12in plus .02in
\centerline{\it 
Department of Physics, University of Southampton,}
\centerline{\it Highfield, Southampton SO17 1BJ, UK}
\vskip .7in plus .35in

\centerline{\bf Abstract}
\smallskip 
A manifestly gauge invariant continuous renormalization group 
flow equation
is constructed for pure $SU(N)$ gauge theory.
The formulation makes
sense without gauge fixing and manifestly gauge invariant
calculations may thus be carried out. 
The flow equation is naturally expressed in terms of fluctuating Wilson loops,
with the effective action appearing as an integral over a `gas' of  Wilson loops.
At infinite $N$, the effective action collapses 
to a path integral over the trajectory of a single particle describing
one Wilson loop. We show that
further regularization of these flow equations is 
needed. (This is introduced in part II.)

\vskip -1.5cm
\Date{\vbox{
{hep-th/9910058}
\vskip2pt{July, 1999.}
}
}

\def\ins#1#2#3{\hskip #1cm \hbox{#3}\hskip #2cm}
\def\etc{{\it etc.}\ }
\def\ie{{\it i.e.}\ }
\def\eg{{\it e.g.}\ }
\def\cf{{\it cf.}\ }

\def\vv{{\it vice versa}}
\def\aka{{\it a.k.a.}\ }

\def\nonp{non-perturbative}
\def\phi{\varphi}
\def\D{{\cal D}}
\def\p{{ p}}
\def\q{{ q}}
\def\r{{ r}}
\def\s{{ s}}

\def\k{{ k}}

\def\x{{ x}}
\def\y{{ y}}
\def\z{{ z}}

\def\O{{\cal O}}

\def\Z{{\cal Z}}
\def\Cu{{\cal C}}

\def\ph#1{\phantom{#1}}
\def\hS{{\hat S}}

\def\ker#1{\!\cdot\!#1\!\cdot\!}

\def\tr{{\rm tr}\,}
\def\la{\lambda}
\def\si{\sigma}

\parindent=15pt

\newsec{Introduction and motivation}
Our main motivation for the present work is to obtain an elegant
gauge invariant Wilsonian renormalization 
group  \ref\kogwil{K. Wilson and J. Kogut, Phys. Rep. 12C (1974) 75.}
 framework formulated directly in the continuum,
as a first step for non-perturbative analytic approximation
methods. Quite generally such methods can prove
powerful,\foot{See for example the reviews 
\ref\algw{D.U. Jungnickel and C. Wetterich,
in {\it The Exact Renormalization Group}, Eds Krasnitz {\it et al},
World Sci (1999) 41, and hep-ph/9902316.}\nref\alg{T.R. Morris, 
in {\it The Exact Renormalization Group}, Eds Krasnitz {\it et al},
World Sci (1999) 1, and hep-th/9810104 (this is updated to conform to 
present notation).}\nref\YKIS{T.R. Morris, 
in {\it Yukawa International Seminar '97}, 
Prog. Theor. Phys. Suppl. 131 (1998) 395, hep-th/9802039.}--\nref\Zak{T.R. Morris, in {\it New Developments 
in Quantum Field Theory},
NATO ASI series 366, (Plenum Press, 1998), 
hep-th/9709100.}\ref\RG{T.R. Morris, 
in {\it RG96}, Int. J. Mod. Phys. B12 (1998) 1343, hep-th/9610012.}. }
and of course there is a clear need for a better \nonp\ understanding
of  gauge theory.
However, several other issues are naturally resolved in the process
of solving this first step. 

In recent years there has been 
substantial progress in solving supersymmetric gauge 
theories \ref\sw{See for example the reviews:
K. Intriligator and N. Seiberg, Nucl. Phys. Proc. Suppl.
45BC (1996) 1, hep-th/9509066;\ C. Gomez and R. Hernandez, 
hep-th/9510023;\ L. Alvarez-Gaum\'e and S.F. Hassan, Fortsch. Phys. 45 (1997)
159, hep-th/9701069.}. These methods involve 
computing a low energy gauge invariant Wilsonian 
effective action, which however, because of the lack of a suitable
framework, is never precisely defined.
Whilst we concentrate here solely on pure Yang-Mills theory, we
see no essential difficulty in generalising the flow equations
to include fermions and scalars and indeed spacetime supersymmetry.
It is clear then that our framework can
underpin these ideas \sw\ref\yosh{S. Arnone, 
C. Fusi and K. Yoshida, J. High Energy Physics 02 (1999) 022.}.

Our framework is here formulated for an $SU(N)$ gauge group, and
is a concise exposition of that reported in ref. \alg (see also
\ref\ymii{T.R. Morris, part II in preparation}).
All of the ideas presented here adapt to other gauge
groups, with minor alterations.\foot{arising from
adapting the completeness relation
for the generators}

Whilst it is an interesting academic issue to establish the existence of
a gauge invariant Wilsonian effective 
action and corresponding flow equation,
its use would ultimately be limited without powerful \nonp\ approximation 
schemes.  Fortunately, a beautiful approximation scheme
lies waiting to be developed: namely 
the large $N$ limit where the gauge group is \eg $SU(N)$ 
\ref\thooft{G. 't Hooft, Nucl. Phys. B72 (1974) 461.}\nref\mig{A. Migdal, 
Ann. Phys. 109 (1977) 365;\
Yu.M. Makeenko and A.A. Migdal, Nucl. Phys. B188 
(1981) 269.}\nref\LN{A.A. Migdal, Phys. Rep. 102 (1983) 199.}\nref\polbo{
A.M. Polyakov, {\sl Gauge fields and Strings}
(Harwood, 1987).}--\ref\others{V.A. Kazakov and I.K. Kostov, Nucl. Phys.
B176 (1980) 199; V.A. Kazakov, Nucl. Phys. B179 (1981) 283.}. 
Typically, the starting point for these methods has been  
Dyson-Schwinger equations for Wilson loops derived at the bare level.
Progress has been hampered by the lack
of corresponding renormalised equations \polbo.
One of the most attractive features of the 
exact renormalization group (RG) is the fact that solutions may readily be
expressed directly in renormalised terms \YKIS. Thus
combining these two approaches removes one obstacle
to solving the large $N$ limit. 

We will see that the large $N$ limit of the Wilson flow equations
however results in an intriguing picture which is (somehow) dual to the
Dyson-Schwinger approach. For example, in this picture the gauge fields
appear not integrated over but take the r\^ole of background field
`spectators' while the Wilson loop, which is fixed (but selectable) in
the Dyson-Schwinger approach, is now dynamical and integrated over. The
analogue of Migdal's observation\mig\
that the large $N$ limit may be expressed in terms of the
expectation value of a single Wilson loop\foot{due to correlators of
Wilson loops decoupling in the planar limit \thooft} 
is here reflected in the fact that the
continuum limit of the Wilsonian effective action, which at finite $N$
may be written in terms of (infinite) sums of integrals of products of
Wilson loops, at
infinite $N$ reduces to a single integral over configurations of just
one Wilson loop. The flow equations reduce to equations determining the
path integral measure for this Wilson loop.  Operators, which may be
viewed as perturbations of this action, also take the form of averages
over Wilson loop configurations. These are nothing but the continuum
counterparts of the `interpolating' operators used in lattice gauge
theory to create propagating glueball states and study their
wavefunctions.  This picture is thus ideally suited to describing both
the gauge fields and the low energy (\eg bound state)
degrees of freedom.

Let us emphasise that the solution of the large $N$ limit in this way,
collapses the quantum field theory to a form
of {\sl single particle quantum
mechanics}. By far the most exciting possibility raised by this
viewpoint, in our
opinion, is that it may finally open the door to the solution of
large $N$ gauge theory. (Quite apart from the obvious theoretical
attractions, the large $N$ limit is expected to be accurate 
in practical situations \eg 10\% accuracy is expected
for many quantities in
$SU(3)$ Yang-Mills \ref\acc{See the review by A.V. Manohar, 
in {\sl PANIC 96}, hep-ph/9607484;\ M.J. Teper, hep-lat/9804008.}.)
In the present paper, this picture lies just below the surface.
We will leave to future work a fuller investigation of this
description, and the simplifications at large $N$
that ensue.
Nevertheless, this picture was
central in guiding us to the construction of 
a consistent flow equation.

A particular problem that has to be faced in this direction,
is that a gauge invariant
effective cutoff function, similarly to gauge invariant higher derivative
regularisation, is not sufficient to regulate all ultra-violet 
divergences. One loop divergences slip through \ref\oneslip{A.A. Slavnov,
Theor. Math. Phys. 13 (1972) 1064;\ B.W. Lee and J. Zinn-Justin,
Phys. Rev. D5 (1972) 3121.}. We cure this problem in the sequel 
by adapting the equations to a novel spontaneously broken supersymmetric
gauge theory, in which the heavy partners play the r\^ole of Pauli-Villars
regulator fields \ymii. In this way all the attractive aspects of the
present framework are preserved while curing this one remaining problem.

The most attractive feature is surely the property that
gauge invariance is explicitly maintained 
at all stages (no gauge fixing or BRST ghosts are required),
resulting in  elegant and highly constrained relations.
One important consequence is that there is no wavefunction renormalization.
The only quantity requiring renormalization is the coupling constant!

Actually, honest non-perturbative approaches to non-Abelian gauge theory
that proceed by gauge fixing, must face up to the challenging problem
of Gribov copies \ref\Grib{V. Gribov, Nucl. Phys. B139 (1978) 1;\
I. Singer, Comm. Math. Phys. 60 (1978) 7;\
See \eg C. Becchi, hep-th/9607181;\ 
P. van Baal, hep-th/9711070;\
M. Asorey and F. Falceto, Ann. Phys. 196 (1989) 209;\
K. Fujikawa, Nucl. Phys. B468 (1996) 355.}. 
Here, these problems are entirely avoided.

In previous exact RG approaches to
gauge theory the authors have gauge fixed, and also 
allowed the effective cutoff to {\sl break} the gauge invariance.
They then seek to recover it in the
limit that the cutoff is removed \ref\qap{M. D'Attanasio and 
T.R. Morris, Phys. Lett. B378 (1996) 213.}\ref\typeai{
C. Becchi, in {\sl Elementary Particles, Field Theory and Statistical
Mechanics}, (Parma, 1993), hep-th/9607188;\
M. Bonini {\it et al}, Nucl. Phys. B409 (1993) 441, B418 (1994) 81, B421 (1994)
81, B437 (1995) 163;\
U. Ellwanger, Phys. Lett. 335B (1994) 364;\
U. Ellwanger {\it et al}, Z. Phys. C69 (1996) 687;\
M. Reuter and C. Wetterich, Nucl. Phys. B417 (1994) 181, B427 (1994) 291;\
K-I. Aoki {\it et al}, Prog. Theor. Phys. 97 (1997) 479, hep-th/9908042,
hep-th/9908043;\ K-I. Kubota and H. Terao, hep-th/9908062;\
M. Pernici {\it et al}, Nucl. Phys. B520 (1998) 469;\
D.F. Litim and J.M. Pawlowski, Phys. Lett. B435 (1998) 181;\ 
M. Simionato, hep-th/9809004.}. As we have indicated,
the present development 
follows a very different route.
In sec. 2, we review Polchinski's
form of Wilson's exact RG molding it into a form suitable
for such a generalisation.
In sec. 3, we then generalise this to $U(1)$
gauge fields as a stepping stone to the full non-Abelian generalisation.
We introduce here several important properties of these
generalisations, namely that gauge invariance may be exactly preserved,
that solutions may be found without gauge fixing, and that 
non-trivial generalisations
exist which leave the partition function invariant and
continue to correspond to
integrating out. In sec. 4, we introduce the full non-Abelian generalisation,
demonstrate that the  gauge field cannot renormalize, introduce the
concept of a `wine', coincident line identities, the definition of the 
coupling $g$,  and discuss some qualitative
criteria satisfied by the formulation namely, 
 `quasilocality', `ultralocality',  and
`integrating out'.
In sec. 5, we develop the gauge field expansion in terms
of traces and products of traces, introducing the effective vertices,
and demonstrate that the large $N$ limit of the effective action has
only a single trace. In sec. 6, we introduce 
and develop the Wilson loop representation which we deliberately leave
this late in the paper to emphasise that the formalism stands separately
from this interpretation, although 
we believe this enables a powerful intuition. The large $N$ limit of the
effective action is then seen to correspond to a form of quantum mechanics
for a single particle and the flow equation is seen to determine the
measure over the fluctuating Wilson loops. In sec. 7, we 
use this interpretation to rapidly develop some of the
general properties of the vertices: the trivial Ward identities
expressing exact preservation of gauge invariance, charge conjugation
invariance, Lorentz invariance, and the coincident line identities.
Sec. 8 sets out the perturbative expansion, demonstrating that the Wilson
loop diagrams are also Feynman diagrams. In this section, we solve
for the classical two, three and four point functions, along the way
showing explicitly how manifestly gauge invariant perturbative
computations may be performed, and explaining how the resulting formulae
are highly constrained by gauge invariance considerations,
in fact to the underlying
Wilson loop picture. We then explain how the $\beta$ function is determined
and give the relevant one-loop contribution. We analyse this and show
how this is at least quadratically divergent (in four dimensions) 
whatever the choice of
cutoff function. This causes subtleties with gauge invariance due
to momentum integral surface terms that do not vanish. We also show
that the coefficients of some of the divergences are not 
even polynomial in the momenta. Appendix A provides
a large momentum analysis of the vertices, necessary to draw these 
conclusions. Along the way we demonstrate that this bad ultraviolet behaviour
is a necessary consequence of the exact preservation of gauge invariance
(and briefly discuss the colinear and small momentum regimes which are 
even more highly constrained by gauge invariance).
Nevertheless we will see in part II  \ymii, that these ultraviolet problems 
can be cured, as already indicated above.

\newsec{The Polchinski equation}

We work in $D$ Euclidean dimensions.
For two functions $f(\x)$ and $g(\y)$ and a 
momentum space kernel $W(p^2/\Lambda^2)$, where
$\Lambda$ is the effective cutoff, we introduce the shorthand:
\eqna\kdef
$$\eqalignno{f\ker{W}g &:=
\int\!\!\!\!\int\!\!d^D\!x\,d^D\!y\
f(\x)\, W_{\x\y}\,g(\y)\quad, &\kdef a\cr
{\rm where}\qquad W_{\x\y} &\equiv\int\!\!{d^D\!p\over(2\pi)^D}
\,W(p^2/\Lambda^2)\,{\rm e}^{i\p.(\x-\y)}\quad.\qquad &\kdef b\cr}$$
Polchinski's \ref\Pol{J. 
Polchinski, Nucl. Phys. B231 (1984) 269.} version of Wilson's exact RG
\kogwil, for the effective interaction 
of a scalar field $S^{int}[\phi]$, may then be written
\eqn\polint{\Lambda{\partial\over\partial\Lambda}S^{int}
=-{1\over\Lambda^2}
 {\delta S\over\delta\phi}^{int}\!\!\!\!\!\ker{c'}
{\delta S\over\delta\phi}^{int}\!\!\!+{1\over\Lambda^2}{\delta 
\over\delta\phi}\ker{c'}{\delta S\over\delta\phi}^{int}\quad.}
Here
$c(p^2/\Lambda^2)>0$ is a {\it smooth}, \ie infinitely differentiable,
ultra-violet cutoff profile,
and prime denotes differentiation with respect to its argument.
The cutoff, which modifies propagators $1/p^2$ to $c/p^2$, 
satisfies $c(0)=1$ so that low energies are unaltered,
and $c(p^2/\Lambda^2)\to0$ as $p^2/\Lambda^2\to\infty$
sufficiently fast that all Feynman diagrams are ultraviolet regulated.
We may write the regularised kinetic term (\ie the Gaussian fixed point)
as 
\eqn\hSs{\hS=\half\,\partial_\mu\phi\ker{c^{-1}}\partial_\mu\phi\quad.}
In terms of the total effective action
$S[\phi]=\hS+\,S^{int}$, and $\Sigma_1:=S-2\hS$,
the exact RG equation reads
\eqn\pol{\Lambda{\partial\over\partial\Lambda}S
=-{1\over\Lambda^2}
 {\delta S\over\delta\phi}\ker{c'}
{\delta\Sigma_1\over\delta\phi}+{1\over\Lambda^2}{\delta 
\over\delta\phi}\ker{c'}{\delta\Sigma_1\over\delta\phi} }
(up to a vacuum energy term that was discarded in \polint\ \Pol). 
The flow in $S$ may be shown directly to correspond to integrating out
higher energy modes
\kogwil\YKIS\ref\wegho{F.J. Wegner and A. Houghton, Phys. Rev. 
A8 (1973) 401.}--\nref\zin{J. Zinn-Justin, 
            ``Quantum Field Theory and Critical Phenomena'' (1993)
             Clarendon Press, Oxford.}\nref\erg{T.R. Morris, 
Int. J. Mod. Phys. A9 (1994) 2411.}\ref\bon{M. Salmhofer, 
Nucl. Phys. B (Proc. Suppl.) {30} (1993), 81.;\
C. Wetterich, Phys. Lett. {B301} (1993), 90;\
M. Bonini {\it et al}, Nucl. Phys. {B418} (1994), 81.},
while leaving the partition function
$\Z= \int\!\!\D\phi\ \e{-S}$ invariant. (For our purposes
we may absorb all source terms into $S$ as spacetime dependent couplings.)
Indeed the invariance of $\Z$ follows from \pol\ because
\eqn\fles{\Lambda{\partial\over\partial\Lambda}\,\e{-S}=
-{1\over\Lambda^2}{\delta\over\delta\phi}\ker{c'}\left(
{\delta\Sigma_1\over\delta\phi}\,\e{-S}\right)}
is a total functional derivative.

\newsec{A flow equation for Abelian gauge fields}

We motivate the form of our non-Abelian exact
RG by first developing an Abelian version.
For $U(1)$ gauge theory, we write the covariant derivative as
$D_\mu=\partial_\mu-ig A_\mu$, where $g$ is the  
coupling.
Clearly if we replace the Gaussian fixed point solution \hSs\ by
the gauge invariant
\eqn\hSui{\hS={1\over4}F_{\mu\nu}\ker{c^{-1}}F_{\mu\nu}}
(where $F_{\mu\nu}=\partial_\mu A_\nu-\partial_\nu A_\mu$), then the
flow equation
\eqn\polui{\Lambda{\partial\over\partial\Lambda}S
=-{1\over\Lambda^2}
 {\delta S\over\delta A_\mu}\ker{c'}
{\delta\Sigma_1\over\delta A_\mu}+{1\over\Lambda^2}{\delta 
\over\delta A_\mu}\ker{c'}{\delta\Sigma_1\over\delta A_\mu}}
is also gauge invariant, while still preserving the invariance of
the partition function under the flow (by  \fles,
replacing $\phi$ by $A_\mu$).
Furthermore, by utilising ref.\ref\ui{T.R. Morris, 
Phys. Lett. B357 (1995) 225.} this may readily be shown
to correspond directly to integrating out higher energy modes \alg.

Note that no gauge fixing has taken place, nor is it necessary for
finding solutions to \polui. 
This exact preservation of gauge invariance will have the important 
consequence,
when generalised to non-Abelian gauge theory, that $D_\mu$ cannot 
renormalise. It is thus convenient to redefine $D_\mu=\partial_\mu-i A_\mu$, 
by changing variables $A_\mu\mapsto A_\mu/g$, since
$A_\mu$ will then not suffer wavefunction renormalisation and only the
coupling  $g$  will renormalise. If we also redefine $\hS\mapsto \hS/g^2$ 
(\ie keep definition \hSui\ for the new fields), \polui\ becomes
\eqn\poluii{\Lambda{\partial\over\partial\Lambda}S +{\beta\over g}
\int\!\!d^D\!x\,A_\mu(\x){\delta S\over\delta A_\mu(\x)}
=-{1\over\Lambda^2}
 {\delta S\over\delta A_\mu}\ker{c'}
{\delta\Sigma_g\over\delta A_\mu}+{1\over\Lambda^2}{\delta 
\over\delta A_\mu}\ker{c'}{\delta\Sigma_g\over\delta A_\mu}\quad,}
where $\beta:=\Lambda \partial g/\partial \Lambda$, and 
$\Sigma_g:=g^2S-2\hS$.
Of course it is still the case 
that $\Lambda{\partial\over\partial\Lambda}\e{-S}$ is a total
functional derivative 
(after addition of a vacuum energy term,
the Jacobian for $A_\mu\mapsto A_\mu/g$).  

This RG equation is not yet in a convenient form for 
generalisation to non-Abelian gauge theory, because the new `$\beta$' term
on the left of \poluii\ is not manifestly gauge 
invariant.\foot{ Gauge invariance follows only once we assume $S$ is
gauge invariant, \ie here $\partial_\mu\left(\delta S/\delta A_\mu\right)=0$.
In the non-Abelian case gauge invariance would not hold separately for
each order in $\hbar$.}
We now come to a central observation of the paper. There are many
other flow equations which have the property that
$\Lambda{\partial\over\partial\Lambda}\,\e{-S}$ is a total
functional derivative, and thus leave 
\eqn\zeff{\Z= \int\!\!\D A\ \e{-S}}
invariant. Furthermore a subset
also correspond to integrating out higher energy modes,
even though direct derivations \kogwil\wegho--\bon\ 
may no longer be possible (\cf ref.\alg\ and the next section). 
In particular, here
we can simply drop this annoying `$\beta$' term! 

Pure $U(1)$ gauge theory is not a good testing ground for these assertions
because the only continuum
solution is $S=\hS/g^2$ with $\beta=0$, \ie the Gaussian fixed point of
free photons. It was thus an
instructive exercise for us to perform a similar analysis for 
 \pol\ and $\lambda\phi^4$ theory, and
demonstrate explicitly that universal
terms, \eg the one-loop $\beta$ function ($D=4$), are unchanged
after mapping $\phi\mapsto\phi/\lambda^{1/4}$ and dropping the 
`$\beta$' term,
and indeed to check that this is true even on more baroque alterations. 
Since the non-Abelian gauge theory exact RG corresponds to our most
important such example, we will not report further on these exercises.

To summarise, for $U(1)$ gauge theory we may write
the exact RG as\eqnn\uiRG\eqnn\Sig
$$\eqalignno{\Lambda{\partial\over\partial\Lambda}S &=-{1\over\Lambda^2}
 {\delta S\over\delta A_\mu}\ker{c'}
{\delta\Sigma_g\over\delta A_\mu}+{1\over\Lambda^2}{\delta 
\over\delta A_\mu}\ker{c'}{\delta\Sigma_g\over\delta A_\mu}\quad, &\uiRG\cr
\ins01{where}\Sigma_g&=g^2S-2\hS\quad, &\Sig\cr}$$
and $\hS$ is given by \hSui. This leaves the partition function \zeff\
invariant:
\eqn\totd{\Lambda{\partial\over\partial\Lambda}\,\e{-S}=
-{1\over\Lambda^2}{\delta\over\delta A_\mu}\ker{c'}\left(
{\delta\Sigma_g\over\delta A_\mu}\,\e{-S}\right)\quad.}
In the next section, we explain why it also corresponds to integrating out 
higher energy modes \alg.

\newsec{A flow equation for non-Abelian gauge fields}

We work with the gauge group $SU(N)$. (These ideas
may easily be extended to general gauge groups).
We are interested in formulating an exact RG for the gauge
field $A_\mu(\x)=A^a_\mu(\x)\tau^a$, the connection for the
covariant derivative $D_\mu=\partial_\mu-iA_\mu$.
The generators $\left(\tau^a\right)^i_{\ph{i}j}$ 
are taken to be Hermitian, in the fundamental representation,
and orthonormalised as ${\rm tr}(\tau^a\tau^b)=\half\delta^{ab}$.

Gauge transformations are of the form
$\delta A_\mu=D_\mu\cdot\omega :=[D_\mu,\omega]$
where $\omega(\x)=\omega^a(\x)\tau^a$. 
Let us stress that exact preservation of this relationship immediately implies
that $A_\mu$ cannot run (and thus has the na\"\i ve unit scaling dimension):
if the gauge field were to suffer multiplicative
wavefunction renormalization by $Z$, we would
have to write $A_\mu\mapsto A_\mu/Z$, destroying the gauge invariance since 
then $\delta A_\mu=(Z-1)\partial_\mu\omega+D_\mu\cdot\omega$. This
argument fails in the gauge fixed theory only because $\omega$ is replaced
by a ghost field in the BRS transformation 
\ref\brst{C. Becchi, A. Rouet and R. Stora, Commun. Math. Phys. 42 (1975) 127.}
leading to pointwise products of fields ($\sim A_\mu\times$
ghost) which are themselves ill defined without further renormalization.

The field strength is 
 $F_{\mu\nu}:=i[D_\mu,D_\nu]$. It will be useful to define
\eqn\funcd{
{\delta\over\delta A_\mu(\x)}:=2\tau^a{\delta\over\delta A_\mu^a}\quad.}
This transforms homogeneously, and its properties can be understood as
follows. To avoid the momentarily extraneous aspects -- 
the $\x$ dependence and $\mu$ index --
let $A^a$ be an adjoint representation 
and write analogously, $\partial/\partial A = 2\tau^a\partial/\partial A^a$.
If $s(A)$ is a function of $A$ such that  
\eqn\var{\delta s(A)=\tr\delta A \,Y\quad,}
then by the completeness relation for $SU(N)$, we effectively isolate
$Y$:
\eqn\varel{{\partial s\over\partial A}=Y-{1\over N}\tr Y\quad.}
This leads, up to $O(1/N)$, to `sowing'
\eqn\sow{\tr X{\partial s\over\partial A} =\tr XY -{1\over N}\tr X\tr Y\quad,}
and `splitting'
\eqn\split{\delta X=Y
\delta A\,Z\qquad\Longrightarrow\qquad\tr{\partial
\over\partial A}X=\tr Y\tr Z -{1\over N}\tr YZ\quad.}

Given a kernel $W(p^2/\Lambda^2)$, 
we construct ${}^{\ph{x}i}_{\x l}\{W\}^k_{j\y}$, the `wine' \alg,
a functional of the kernel which is its
gauge covariantization, 
incorporating parallel transport of the 
tensor representation.  Thus, if
$v^j_k(\y)$ and $u^l_i(\x)$ are two  
$N\otimes{\bar N}$ representations of the gauge
group $SU(N)$, the gauge invariant generalisation of \kdef{} is
\eqn\wc{u\{W\}v:=\int\!\!d^D\!x\,d^D\!y\,
u^l_i(\x)\,{}^{\ph{x}i}_{\x l}\{W\}^k_{j\y}\,v^j_k(\y)\quad,}
where without loss of generality we may insist that $\{W\}$ 
satisfies $u\{W\}v\equiv v\{W\}u$.
The index flow, or parallel transport, is
illustrated in \fig\ifl{Index flow for ... 
${}^{\ph{x}i}_{\x l}\{ \}^k_{j\y}\,$ eqn.\wc.}.
\midinsert
$$
\epsfxsize=0.25\hsize\epsfbox{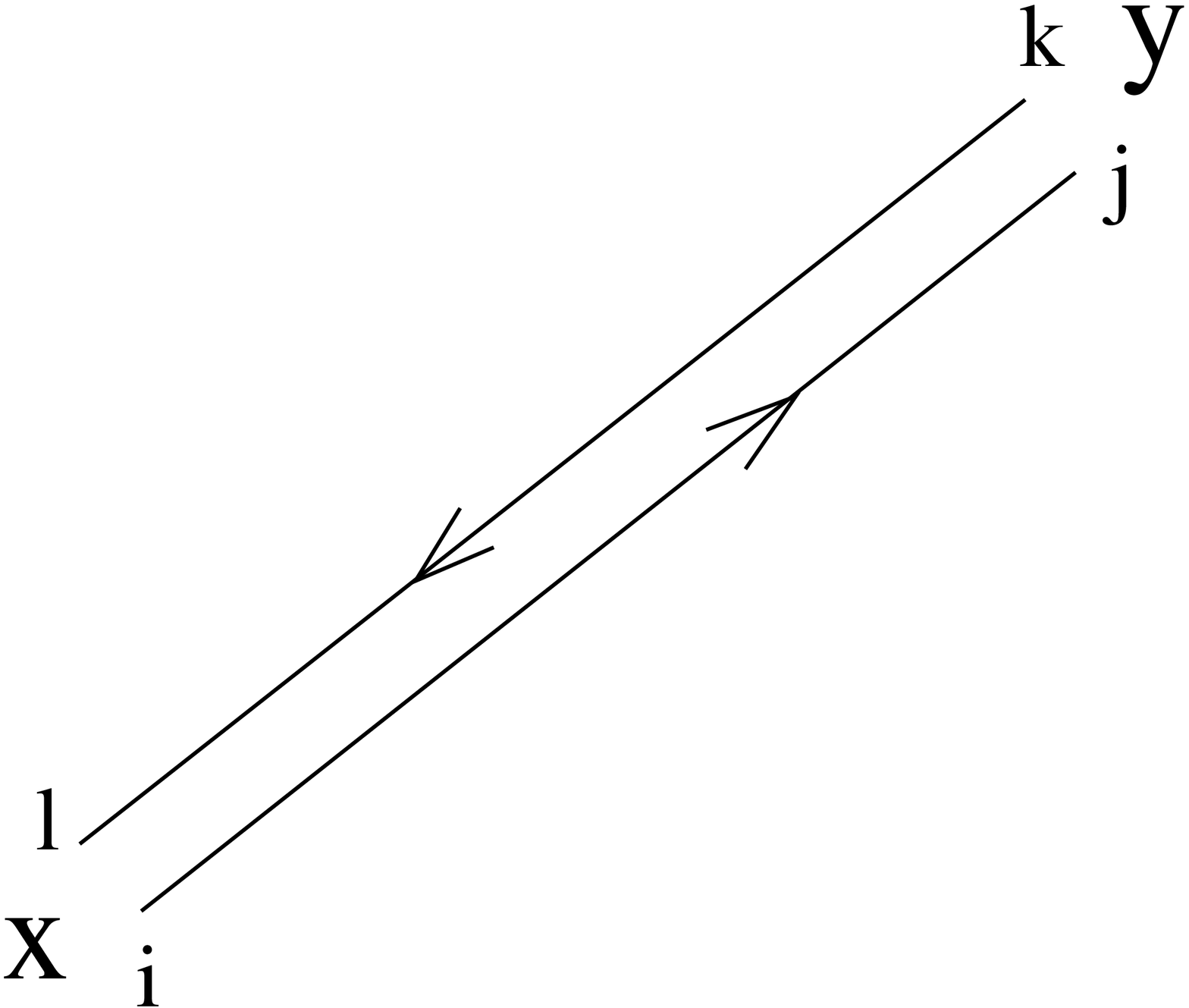}
$$
\centerline{ {\bf Fig.1.} Index flow in eqn.\wc.}
\endinsert
Expanding this in terms of the gauge field $A_\mu$ defines the 
wine-vertices:
\eqnn\wv
$$\displaylines{ 
{}^{\ph{x}i}_{\x l}\{W\}^k_{j\y}= \hfill \wv\cr 
\sum_{m,n=0}^\infty\int\!\!
d^D\!x_1\cdots d^D\!x_n\,d^D\!y_1\cdots d^D\!y_m\,
W_{\mu_1\cdots\mu_n,\nu_1\cdots\nu_m}
(\x_1,\cdots,\x_n;\y_1,\cdots,\y_m;\x,\y) \cr
\hfill [A_{\mu_1}(\x_1)\cdots A_{\mu_n}(\x_n)]^i_{\ph{i}j}\,
[A_{\nu_1}(\y_1)\cdots A_{\nu_m}(\y_m)]^k_{\ph{k}l}\quad.\cr}$$
The $m=0$ case should be interpreted as follows: it has vertices 
$W_{\mu_1\cdots\mu_n,}(\x_1,\cdots,\x_n;;\x,\y)$ 
and the second product of gauge
fields is replaced by $\delta^k_{\ph{k}l}$. The $n=0$ case is defined
similarly. We will write the $m=0$ vertices more compactly as
\eqn\compac{W_{\mu_1\cdots\mu_n}(\x_1,\cdots,\x_n;\x,\y)
\equiv W_{\mu_1\cdots\mu_n,}(\x_1,\cdots,\x_n;;\x,\y)\quad.}
In addition, the $m=n=0$ vertex is just the original 
kernel \kdef b, \ie 
\eqn\mno{W_,(;;\x,\y)\equiv W_{\x\y}\quad.}
This expansion is illustrated in \fig\winexp{Note replacement thick
black lines by thin lines and blobs ...}.
\midinsert
$$
\epsfxsize=\hsize\epsfbox{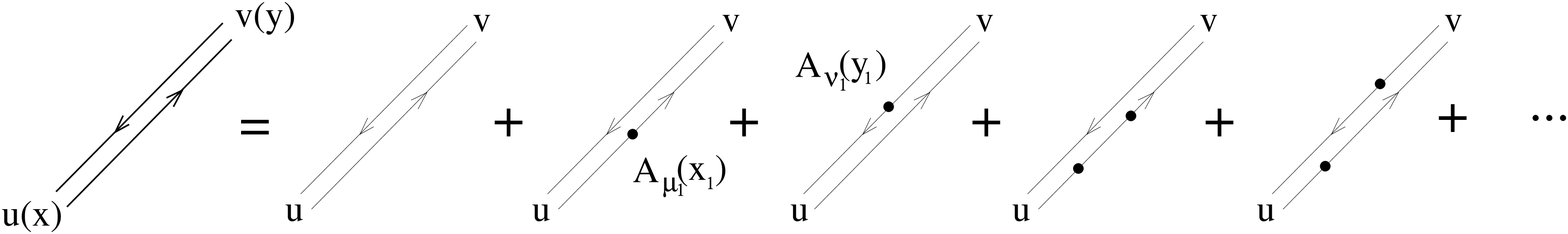}
$$
\centerline{ {\bf Fig.2.} Expansion of the wine in gauge fields. The
thick black lines stand for the full series.}
\endinsert
Up to the requirement that it is still {\it smooth} in momentum space
(\ie that all vertices are Taylor expandable to all orders in momenta) \alg\
and some symmetry constraints (\cf below \wc\ and later),
the covariantization $\{W\}$ is of our choosing. 
However, for simplicity we further impose
some `coincident 
line' identities (described in sec.5) which in particular result in
\eqn\coid{v^j_k(y)=\delta^j_k g(y)\quad \forall y\qquad\Longrightarrow
\qquad
u\{W\}v=({\rm tr}u)\ker{W}g\quad,}
as represented  in \fig\fcoi{coincident line id on p6 Rome,
dotted line represents just the $W$ kernel \kdef b. }.
\midinsert
$$
\epsfxsize=0.4\hsize\epsfbox{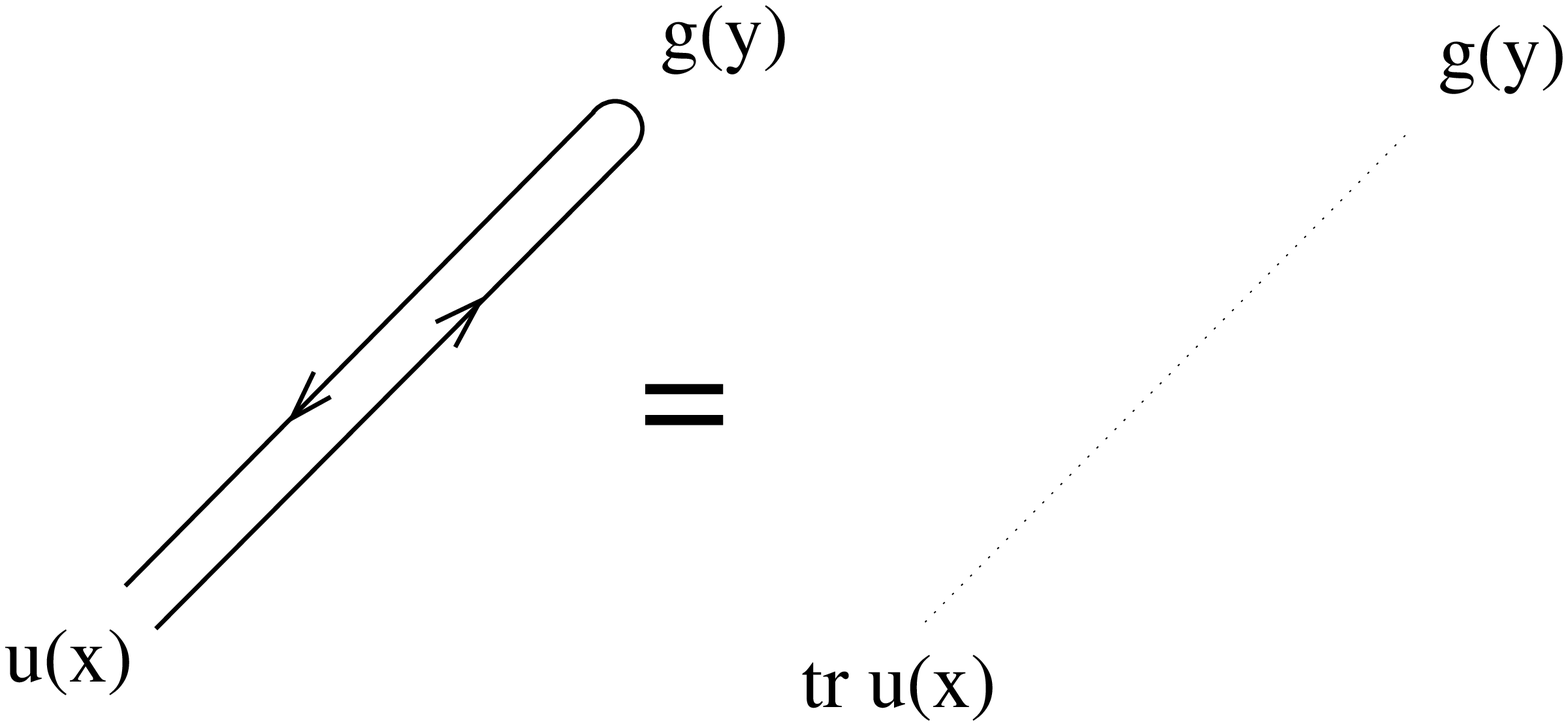}
$$
\centerline{ {\bf Fig.3.} Coincident line identity. The dotted line is
just the $W$ kernel \kdef b.}
\endinsert
For example, we could simply include in \kdef a
two straight Wilson lines 
\eqn\straw{u\{W\}v=\int\!\!\!\!\int\!\!d^D\!x\,d^D\!y\
W_{\x\y}\, \tr u(\x)\Phi[\Cu_{\x\y}]v(\y)\Phi^{-1}[\Cu_{\x\y}]\quad,}
$\Cu_{\x\y}$ being the straight line between $\x$ and $\y$.
(Later we will consider more general curves. Recall that the Wilson line is
the path ordered exponential:
\eqn\defWl{\eqalign{\Phi[\Cu_{\x\y}]&
=P\exp-i\int_{\Cu_{\x\y}}\!\!\!\!\!\!dz^\mu
A_\mu(z)\quad,\cr
&=1-i\int_0^1\!\!\!\!d\tau\,{\dot z}^\mu\! A_\mu(z)
-\int_0^1\!\!\!\!d\tau_2\!\int_0^{\tau_2}\!\!\!\!\!d\tau_1\,
{\dot z}\!\cdot\! A(\tau_1)\,
{\dot z}\!\cdot\! A(\tau_2)\ +\cdots\quad,}}
where we have parametrized $\Cu_{\x\y}$ by $z^\mu(\tau)$, 
$\tau\in[0,1]$, $\z(0)=\x$, $\z(1)=\y$.)
Another choice can be obtained by
utilising the momentum representation, to write
\eqn\ourchoice{u\{W\}v={\rm tr}\int\!\!d^D\!x\, u(\x)
\,W(-D^2/\Lambda^2)\cdot v(\x)}
(where, as above \funcd, the covariant derivatives act by commutation).
We will not need to specify the choice in this paper.

We covariantize \hSui\ to 
\eqn\seed{\hS=\half F_{\mu\nu}\{c^{-1}\}F_{\mu\nu}\quad.}
(Note that gauge invariance forces us to include not just the Gaussian fixed
point but also the interactions.)
Our (insufficiently regularised) exact RG is then just the 
covariantization of \uiRG\ \alg:
\eqn\basRG{\Lambda{\partial\over\partial\Lambda}S=
-{1\over2\Lambda^2}{\delta S\over\delta A_\mu}\{c'\}
{\delta\Sigma_g\over\delta A_\mu}
+{1\over2\Lambda^2}{\delta\over\delta A_\mu}\{c'\}
{\delta\Sigma_g\over\delta A_\mu}\quad.}
Just as in \pol, the first term on the RHS is the {\it classical} term,
yielding the tree corrections, while the second, {\it quantum}, term,
generates the loop corrections.
Note that we have assumed the same functional relationship for the
covariantization of $c^{-1}$ and $c'$ [\eg
as in \ourchoice]. This is convenient but not necessary.
The flow preserves the corresponding  partition
function \zeff, just as before, because \totd\ holds -- on replacing
$\cdot c'\!\cdot$ with $\{c'\}$.

The coupling, $g$, which appears in \basRG\ through 
\Sig, is defined by $S$ via 
its unique derivative-squared term:
\eqn\defg{S={1\over2g^2}\,{\rm tr}\!\int\!\!d^D\!x\, F_{\mu\nu}^2
+O(\partial^3)}
(discarding the vacuum energy). Note that  
$S$ must have a derivative expansion 
to all orders, because $\Lambda$ must play
the r\^ole of an infrared cutoff in its 
vertices \erg\YKIS, as explained below.

One of the great attractions of gauge theory is that in four dimensions,
there are no relevant or marginal 
operators\foot{with respect to the Gaussian fixed point}
we can put in by hand. In this case, $g$
is the only coupling, and furthermore the self-similarity of the continuum solution \YKIS\ ensures that the only explicit
dimensionful parameter is $\Lambda$ whose appearance is then determined
by dimensions. Thus in $D=4$ dimensions we can write for the continuum
solution
\eqn\deforg{S={1\over2g^2}\,{\rm tr}\!\int\!\!d^4\!x\, F_{\mu\nu}^2
+O(\partial^3/\Lambda)}
(again discarding the vacuum energy).  

In preparation for the large $N$ limit we write $g\mapsto g/\sqrt{N}$,
$S\mapsto N S$. This has no effect on \Sig, \defg\ or \deforg,  while \basRG\
becomes
\eqn\RG{\Lambda{\partial\over\partial\Lambda}S=
-{1\over2\Lambda^2}{\delta S\over\delta A_\mu}\{c'\}
{\delta\Sigma_g\over\delta A_\mu}
+{1\over N}{1\over2\Lambda^2}{\delta\over\delta A_\mu}\{c'\}
{\delta\Sigma_g\over\delta A_\mu}\quad.}
As we will show explicitly in the next section,
the quantum term survives the large $N$ limit only in those terms
in which each loop correction is accompanied by ${\rm tr}1=N$  \alg.

We finish this section by discussing some important
qualitative criteria satisfied by \basRG\ or \RG.
Recall that it is a fundamental requirement of the renormalization group
that the Kadanoff
blocking transformation \kogwil\ should affect variables only in a
localised patch; long range interactions should appear only after
infinitely many steps (analogously here $\Lambda\to0$).

On the one hand, obviously, we must therefore require that each RG step
is free from infrared singularities,\foot{Actually
this statement is strictly true only for a smooth (\ie infinitely
differentiable) cutoff. More subtle cases, \eg sharp cutoff
\ref\truncm{T.R. Morris, Nucl. Phys. B458[FS] (1996) 477.}, will not be
considered here.} 
equivalently here that {\it the flow equation must
have an all-orders Taylor expansion in small external momenta}. 
We call this the requirement of `quasilocality' \alg. 
This implies that the same is true of $S$, as we claimed above, 
providing only that we ensure that
$\Lambda$-integration constants are also chosen quasilocal. As we will 
confirm explicitly, 
quasilocality then follows from the smoothness of the wine-vertices and $\hS$.
After integrating down to some intermediate scale $\Lambda$, the long
range interactions are hidden in $S$ in this quasilocality (since the
modes that are integrated out are infrared cutoff by $\Lambda$). 

On the
other hand, we want to ensure that unintended fundamental physics or
even non-local non-physics is not also hidden in this quasilocality.
We call this the requirement of `ultralocality'.  For momenta much larger
than the infrared cutoff $\Lambda$, 
the typical $\sim 1/p^2$ behaviour of fundamental
propagating modes does show up in the vertices of $S$.  
Therefore the
requirement of ultralocality corresponds to ensuring that 
there are no spurious inverse powers of momenta in this
asymptotic expansion \alg, in particular  
that we do not introduce any propagator-like terms into \seed.

(Actually, despite the intuition,
it is not clear to us whether this requirement is really needed. 
A more careful study than that reported in ref. \alg, shows that
it is not in fact possible to generate an arbitrary quasilocal $S^0$ 
from a quasilocal $\hS$, rather there are some restrictions on 
additions to $S^0$. It may be that
the physics is still correct but hidden in a 
change of variables. To definitively answer these issues requires
generalising the recipe for extracting correlators and thus
scattering matrix elements from the effective action \erg\YKIS.)

Finally, we can see indirectly that \basRG\ must still
correspond to integrating out, once this flow equation is completely
regularised, as follows.
In this case, by definition all momentum integrals
are bounded. But since $\Lambda$ appears as the ultraviolet cutoff scale in
these integrals, 
this means that contributions from momenta larger than some scale $q$
must vanish in the limit $q/\Lambda\to\infty$.
Thus we see that as $\Lambda\to0$, the remaining
contribution from any fixed range of
non-vanishing momentum scales, disappears. Since the partition function
\zeff\ is unchanged under the flow, the contributions from a given fixed 
momentum scale must still be in there somewhere, and the 
only other place they can be, is to already be
encoded in the effective action -- \ie the modes have been
integrated out. (Strictly it is the 
partition function \zeff\ that we must show is completely regularised, and
not just the flow equation. It is hard to see here however, how this property
could separately fail.)
 A more intuitive, but incomplete, requirement for
integrating out to be taking place (which
is also satisfied) was discussed in ref. \alg.

\newsec{Gauge field expansion and the large N limit}

The Wilsonian effective action $S$, being gauge invariant, has an expansion
in traces and products of traces:\eqnn\Sex

$$\displaylines{S =\sum_{n=2}^\infty{1\over n}\int\!\!d^D\!x_1\cdots d^D\!x_n\,
S_{\mu_1\cdots\mu_n}(\x_1,\cdots,\x_n)\ 
\tr A_{\mu_1}(\x_1)\cdots A_{\mu_n}(\x_n)\hfill\cr
\ph{S}+{1\over2!}\sum_{m,n=2}^\infty{1\over nm}\int\!\!
d^D\!x_1\cdots d^D\!x_n\,d^D\!y_1\cdots d^D\!y_m\,
S_{\mu_1\cdots\mu_n,\nu_1\cdots\nu_m}(\x_1,\cdots,\x_n;\y_1,\cdots,\y_m)
\hfill\cr
\hfill\tr A_{\mu_1}(\x_1)\cdots A_{\mu_n}(\x_n)\
\tr A_{\nu_1}(\y_1)\cdots A_{\nu_m}(\y_m)\cr
\ph{S}+\cdots\quad.\hfill\Sex\cr}$$
This is represented graphically in \fig\sexp{.. the diags of p9 Rome} 
(\cf \winexp. Note that the combinatorics in the figures are 
those of Feynman rules:
\ie each diagram stands for the sum over all ways of assigning the gauge
fields to the points, whilst respecting the order in the traces.)
\midinsert
$$
\epsfxsize=\hsize\epsfbox{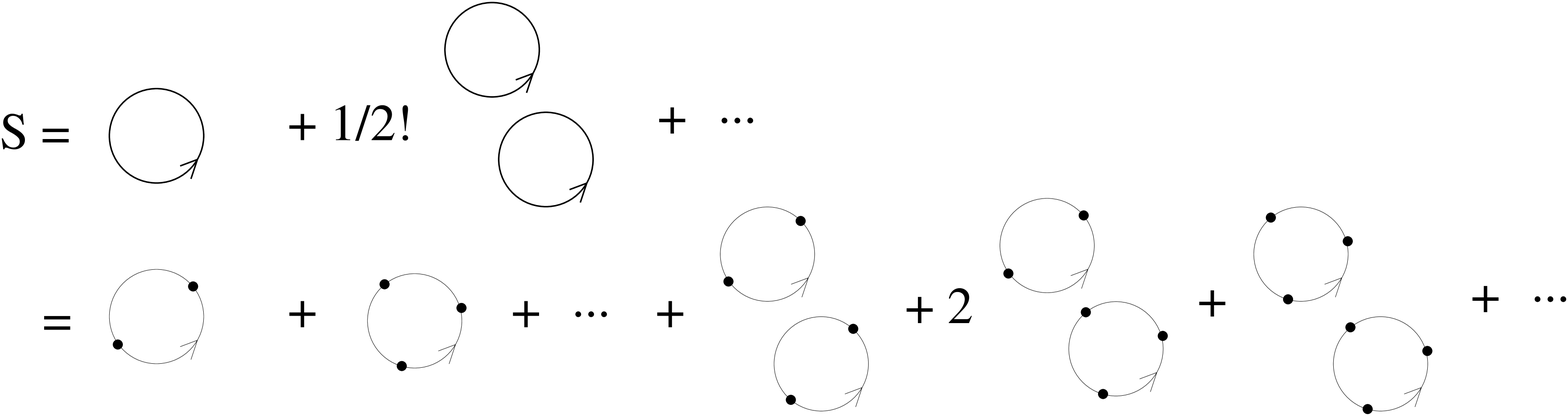}
$$
\centerline{ {\bf Fig.4.} Expansion of the action in traces, and
products of traces, of gauge fields.}
\endinsert
The momentum space vertices are written as
\eqn\ft{ S_{\mu_1\cdots\mu_n}(\p_1,\cdots,\p_n)\
(2\pi)^D\delta( 
\sum_{i=1}^np_i)  
=\int\!\!d^D\!x_1\cdots d^D\!x_n\,\e{-i\sum_i\x_i.\p_i}
S_{\mu_1\cdots\mu_n}(\x_1,\cdots,\x_n)\quad,} 
where all momenta are taken pointing into the vertex, and similarly for
the vertices in \wv. $\hS$ has an expansion in only single
trace vertices $\hS_{\mu_1\cdots\mu_n}(\x_1,\cdots,\x_n)$, as follows
from \seed. In fact, combining \wv\ and \seed, we may easily read off
expressions for these vertices \alg:
\eqn\hSex{\eqalign{\hS_{\mu\nu}(\p) &\equiv\hS_{\mu\nu}(\p,-\p)=
2\Delta_{\mu\nu}(\p)/c_\p\cr
\hS_{\mu\nu\lambda}(\p,\q,\r) &={2\over c_\p}(p_\lambda\delta_{\mu\nu}-
p_\nu\delta_{\lambda\mu})
+2c^{-1}_\nu(\q;\p,\r)(p_\lambda r_\mu-p.r\delta_{\lambda\mu})
+{\rm cycles}\cr 
\hS_{\mu\nu\lambda\sigma}(\p,\q,\r,\s) &={1\over c_{\p+\q}}
(\delta_{\sigma\mu}\delta_{\lambda\nu}-
\delta_{\lambda\mu}\delta_{\nu\sigma})+2c^{-1}_\nu(\q;\p,\r\!+\!\s)
(p_\sigma\delta_{\lambda\mu}-p_\lambda\delta_{\sigma\mu})\cr
&\ph{=}+2c^{-1}_\sigma(\s;\p,\r\!+\!\q)
(p_\nu\delta_{\mu\lambda}-p_\lambda\delta_{\mu\nu})
+2c^{-1}_{\nu\lambda}(\q,\r;\p,\s)(p_\sigma s_\mu-p.s\delta_{\sigma\mu})\cr
&\ph{=}+c^{-1}_{\nu,\si}(\q;\s;\p,\r)(p_\lambda r_\mu-p.r\delta_{\lambda\mu})
+{\rm cycles}\cr
}}
\etc,
where in the two-point vertex we set $\p_1=-\p_2=\p$, and introduce
the shorthand $c_\p\equiv c(p^2/\Lambda^2)$
and the transverse combination
$\Delta_{\mu\nu}(\p) :=p^2\delta_{\mu\nu}-p_\mu p_\nu$, 
in the three-point vertex 
we add the two cyclic permutations of $(p_\mu,q_\nu,r_\lambda)$,
and in the four-point vertex the three cyclic permutations
of $(p_\mu,q_\nu,r_\lambda,s_\sigma)$. 
Note that, since $c_p$ and all the wine-vertices are smooth, all the 
$\hS$-vertices have the same property.  

Concentrating on the single trace terms of $S$ (or $\hS$)
 we see from \wv, \Sex, and
\sow, that $\{c'\}\delta S/\delta A_\mu$ opens up the trace by removing
one $A_\mu$ and replacing it with the ends of the lines of parallel transport
in \ifl. 
In the classical term of \RG\ the leftmost  $\delta/\delta A_\mu$
does the same with another (copy of the) action, while in
the quantum term, the leftmost $\delta/\delta A_\mu$ 
opens up the same trace again and attaches the other ends of the parallel
transport lines, splitting the trace in two, as in \split. 
Diagrammatically, 
the single trace terms of $S$, under a small change in $\Lambda$, 
thus induce
the contributions in
\fig\fflow{Diag of p10 Rome, but include an extra $1/N$ in loop diags.
In here a circumflex in the circle stands for $\hS$, a dotted line for
the attachment of $c'_{\x\y}$ only as in \fcoi,
this term arising from \coid\ and
the $1/N$ `remainder' terms in \varel, \sow\ and \split.}. 
\midinsert 
$$\eqalign{&\Lambda{\partial\over\partial\Lambda}\,
\vcenter{\epsfxsize=0.085\hsize\epsfbox{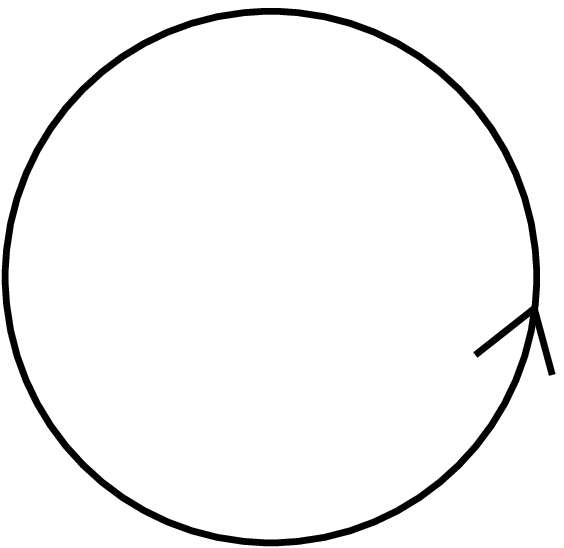}}\,
=\cr
-&g^2\left\{\vcenter{\epsfxsize=0.36\hsize\epsfbox{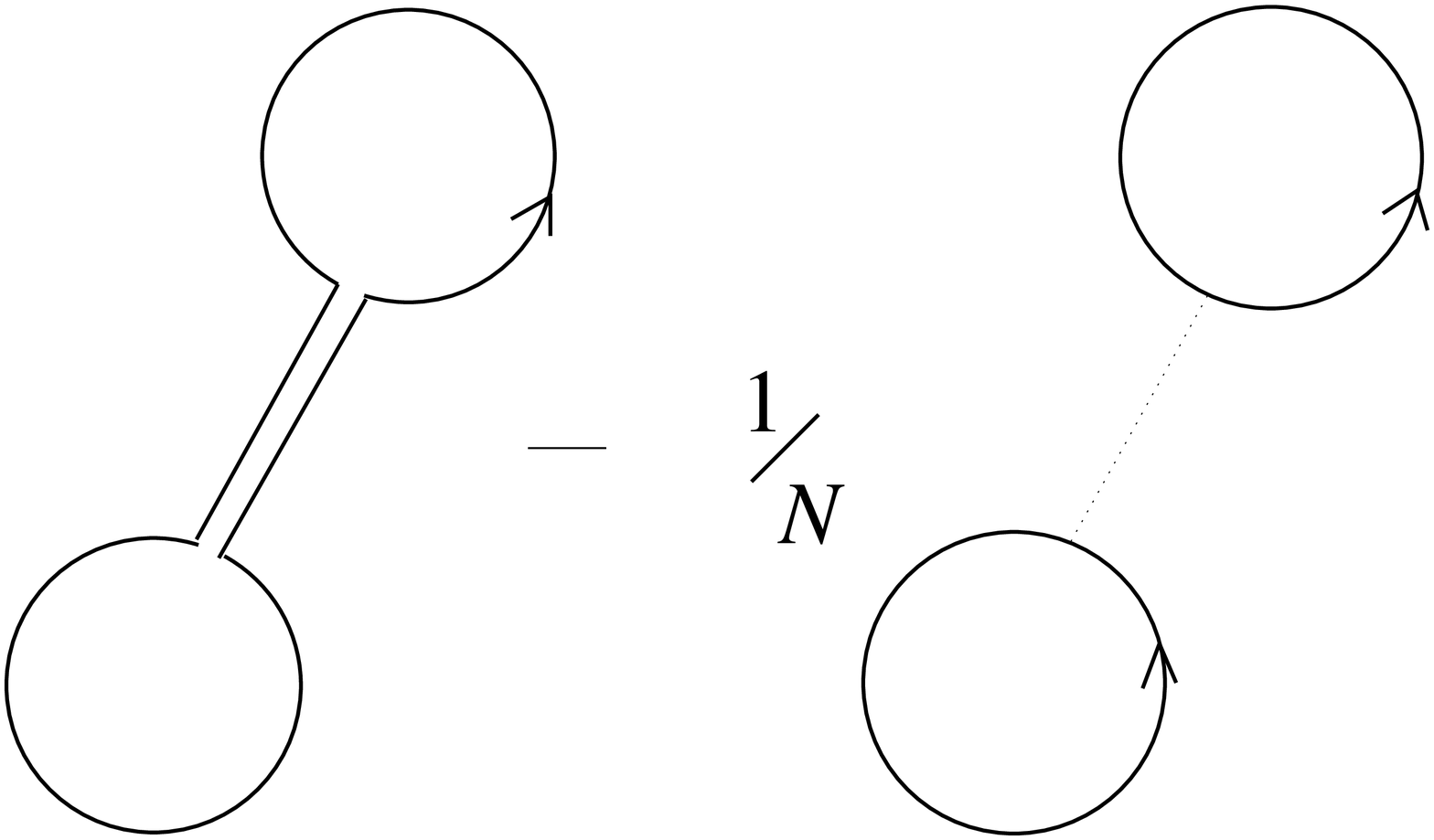}}\right\}
+2\left\{\vcenter{\epsfxsize=0.36\hsize\epsfbox{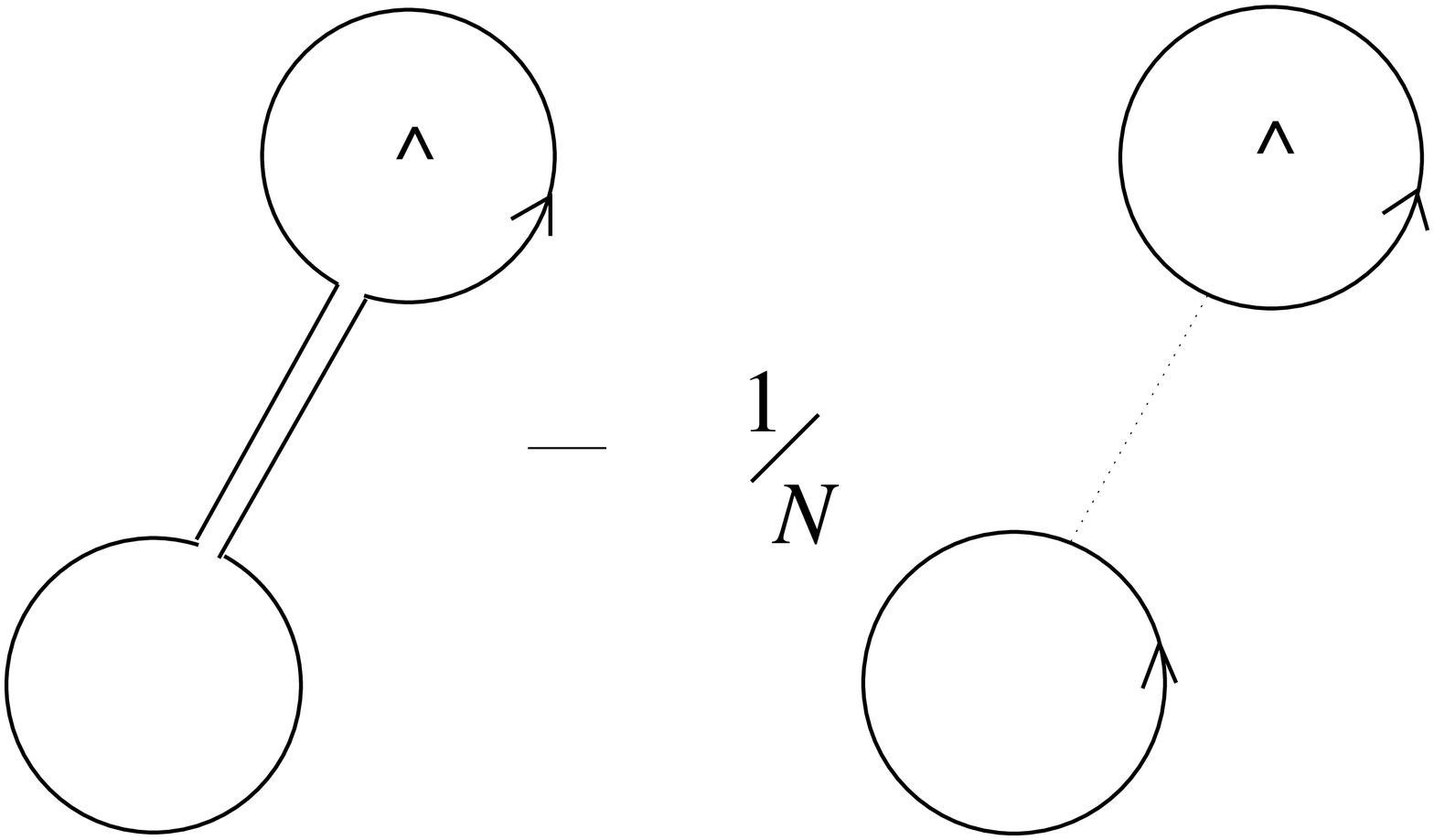}}\right\}\cr
+&{g^2\over N}\left\{
\vcenter{\epsfxsize=0.3\hsize\epsfbox{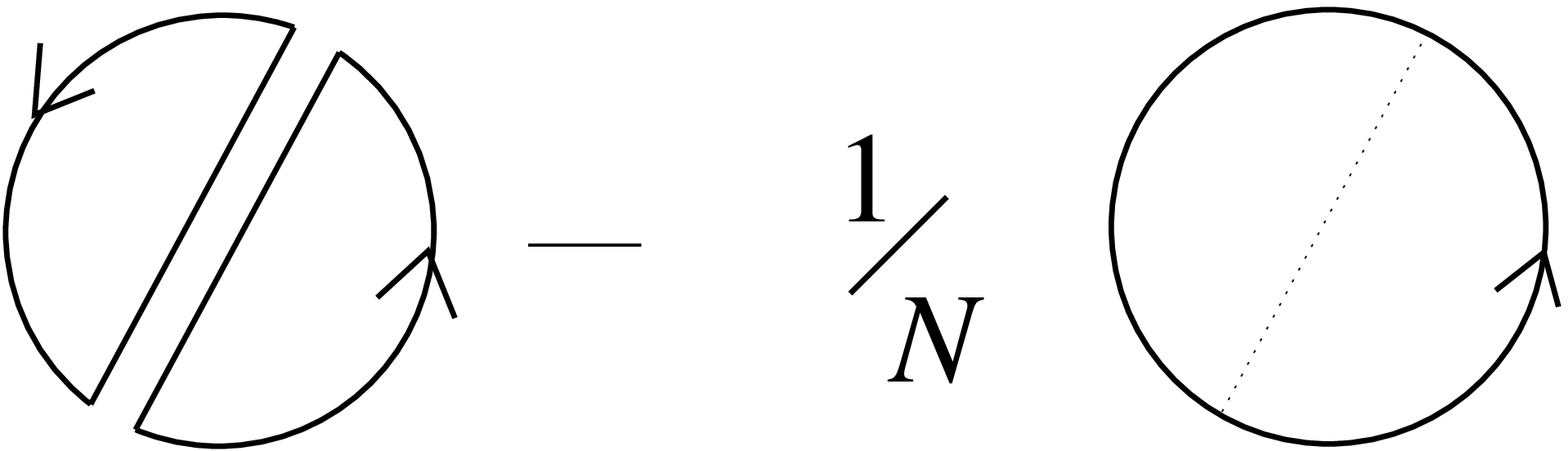}}\right\}
-{2\over N} \left\{\vcenter{\epsfxsize=0.3\hsize\epsfbox{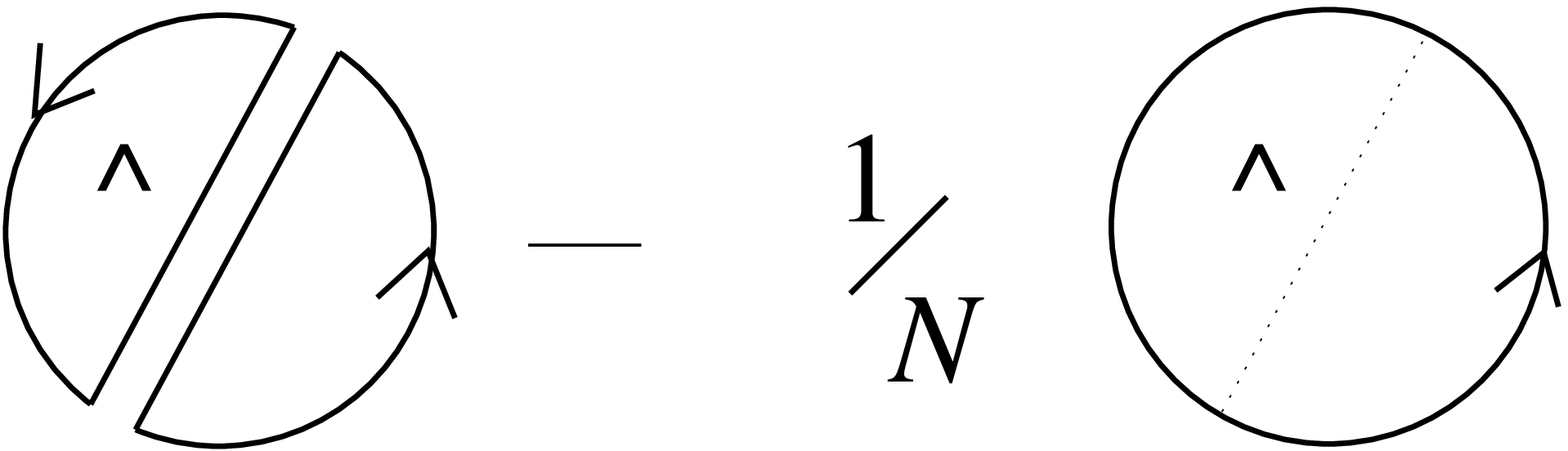}}\right\}
}$$
{\bf Fig.5.} Diagrammatic representation of the flow equation.
A circumflex in the circle stands for $\hS$, a dotted line for attachment
of $c'_{\x\y}$ only, as in \fcoi.
\endinsert
Strictly speaking we should include in the quantum term of \RG, 
the contributions as in
\fig\tail{wine biting tail as in fig.5 Alg} where the leftmost
functional derivative attacks the parallel transport lines themselves.
However, we can avoid them by a limiting procedure keeping the
position of the functional derivative away from the wine \alg.
In fact these terms will vanish once the Pauli-Villars contributions 
are included \ymii.
\midinsert
$$
\epsfxsize=0.28\hsize\epsfbox{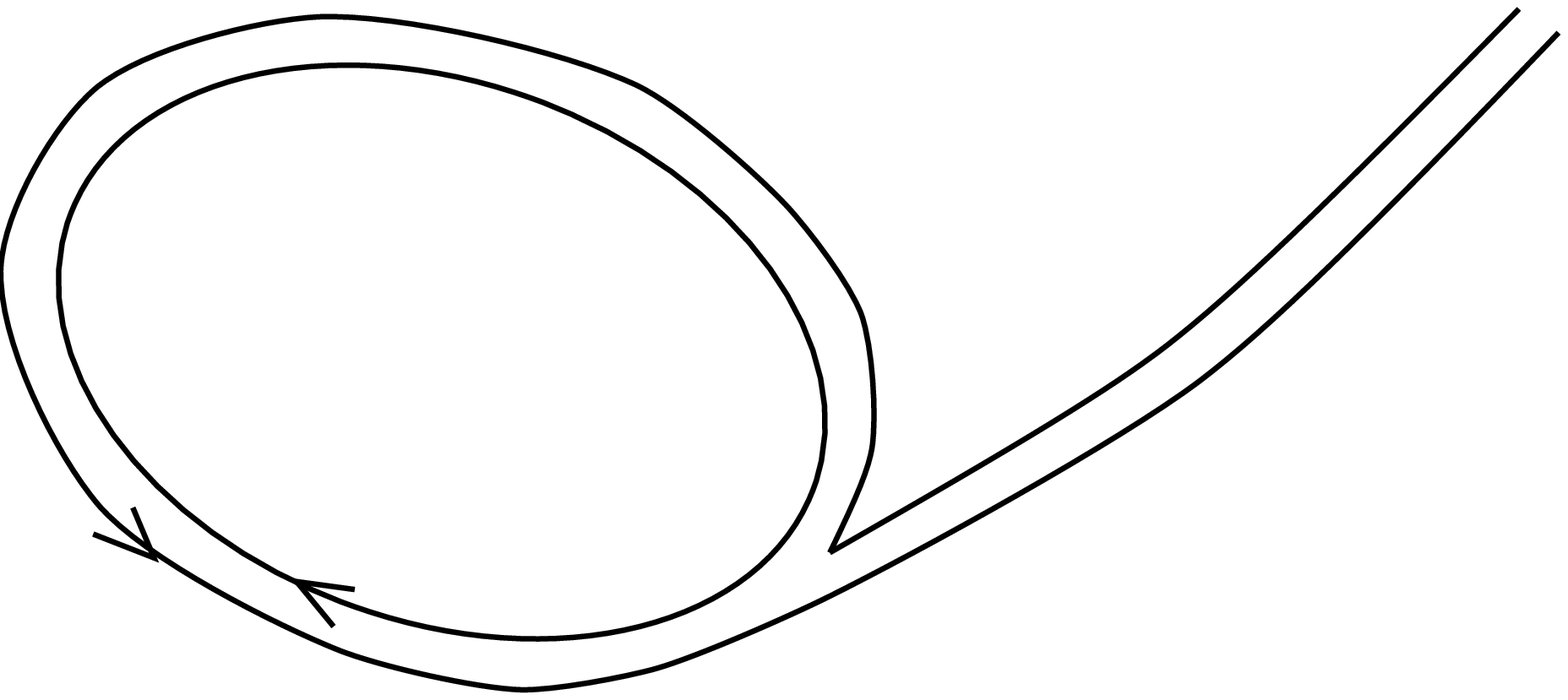}
$$
\centerline{ {\bf Fig.6.} A wine biting its own tail.}
\endinsert

As $\Lambda\to\infty$, $S$ is given by just
the first term in \deforg, and thus is a single trace in this limit.
We see from \fflow\ that those terms with a product of two
traces are down by factors of $1/N$. In the large $N$ limit the only 
quantum terms that survive are the split-open traces where all the
gauge fields lie in only one of the two new traces, as in 
\fig\flargeN{Fig 11 Algarve all gauge fields on one of two traces}. 
Thus in the large $N$ limit,
the single trace  property is preserved by the flow and the effective
action at any $\Lambda$ contains only single trace terms.
\midinsert
$$
\epsfxsize=0.28\hsize\epsfbox{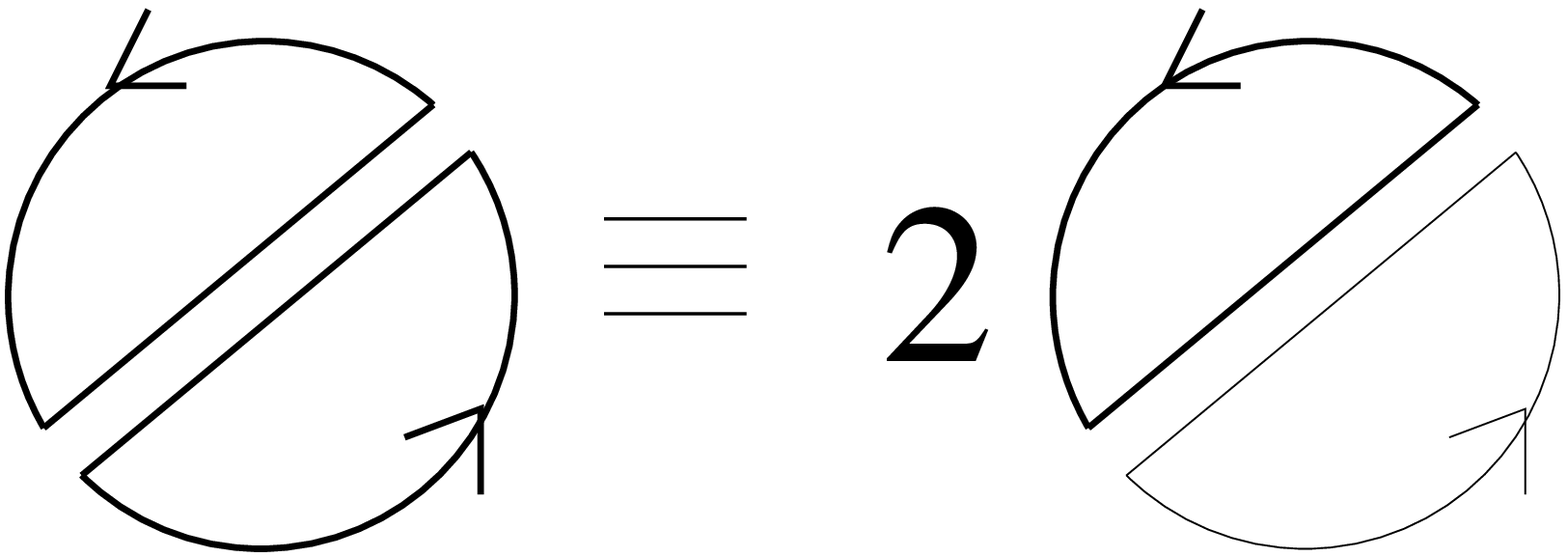}
$$
{\bf Fig.7.} In the large $N$ limit, the surviving terms have all gauge
fields living on one of the two traces, the no-gauge-fields trace 
contributing tr$1=N$.
\endinsert

However, whilst it is true that the large $N$ limit of the effective
action is a single trace, the limit of the flow equation itself may be
more subtle: two-trace terms with just two gauge fields in one of the
traces are also formally down by a factor of $1/N$, but the attachment of 
a field-free
wine 
to these two gauge fields in a higher order quantum correction,
will then yield a further factor of $1/N$ (for the quantum correction)
and a factor of $N^2$ for two field-free traces,  thus contributing
in principle a finite amount in the $N\to\infty$ limit. To one loop this
does not happen because  two-trace terms
with just two fields in one of the traces, vanish at tree level
for symmetry reasons, as we prove in sec. 8. We leave to the future, 
the question of whether or not this more subtle large $N$
limit of the flow equations takes place at higher loop order.

\newsec{Wilson loop representation}

The Wilson loop representation was
central to our exposition in \alg, and indeed plays a powerful
r\^ole in our own thinking, but we have intentionally
avoided introducing it until this point
to emphasise that the flow
equations and their properties hold {\sl entirely separately from this
representation}, even though we find that these properties show
up most clearly through this representation.

We have already mentioned, \cf \straw,
that we could covariantize the kernel by using straight
Wilson lines. More generally, we could use curves $\Cu_{xy}$,
although to ensure
Lorentz invariance these would have to be averaged over their orientations.
The most general covariantization, preserving the fact that in
\wv\ the gauge fields appear in just two strings,  is then given by 
 (path) integrals  and/or sums  over all the configurations of these
two curves, with a normalised
measure of our own choosing [up to the Lorentz covariance requirement above, 
the smoothness requirement above \coid, and exchange symmetry below \wc]. However, we simplify this
by traversing back along a coincident Wilson line, \ie
back from $y$ to $x$ along the same curve: 
\eqn\wili{
u\{W\}v=\int\!\!\!\!\int\!\!d^D\!x\,d^D\!y\int\!\!\D\Cu_{\x\y}\
\tr u(\x)\Phi[\Cu_{\x\y}]v(\y)\Phi^{-1}[\Cu_{\x\y}]\quad.}
The measure (hidden in the definition of $\D\Cu_{xy}$) 
does not of course
depend on the parametrization of the path, and is normalised by
\eqn\pnorm{\int\!\!\D\Cu_{\x\y}\ 1 = W_{\x\y}\quad,}
as follows from \mno. (As a  trivial
example, in the case of \straw, it has support
only on the straight line $\Cu_{xy}$ where the measure collapses to
multiplication by $W_{xy}$.) Clearly \wili\ and \pnorm\ imply 
the identity \coid\ we introduced earlier.

Similarly the most general form for the action is in terms of  path
integrals over Wilson loops $\phi[\Cu]={\rm tr}\,\Phi[\Cu_{\x\x}]$
($\Cu$ being a closed contour, $\x$ some arbitrary point on it, and
$\Cu_{\x\x}$ being the marked contour starting and finishing at $\x$).
The most general single trace term is obtained from an integral over all
configurations of a single Wilson loop, the most general two-trace terms
from two Wilson loops and so on. Thus \Sex\ may be represented as
\eqn\defWS{S=\int\!\!\D\Cu\,\phi[\Cu]+{1\over2!}\int\!\!\D[\Cu_1,\Cu_2]\,
\phi[\Cu_1]\,\phi[\Cu_2]\ +\cdots\quad,}
and $\hS$ may be written
\eqn\defWhS{\hS=\int\!\!\D{\hat\Cu}\,\phi[{\hat\Cu}]}
[by \hSui]. While the measure $\D{\hat\Cu}$ is of our choosing, 
the measures $\D\Cu$,
$\D[\Cu_1,\Cu_2]$, \etc, are determined by the flow equations. 

By virtue of the fact that 
$$\delta\Phi[\Cu_{xy}]=-i\int_{\Cu_{\x\y}}\!\!\!\!\!\!dz^\mu\
\Phi[\Cu_{xz}]\,\delta A_\mu(z)\, \Phi[\Cu_{zy}]\quad,$$
where $\Cu_{xz}$ ($\Cu_{zy}$) is the part of $\Cu_{xy}$
before (after) $z$, all our previous figures still hold but
as `snapshots' of the
Wilson loops and/or lines. We only have to remember that wherever the
`wine' attaches, it is integrated around the curve by 
$-i\int dz^\mu$. As a result, our previous figures, \eg \fflow,
are no longer just `hieroglyphics' --indicating the form of the resulting 
traces -- but come `alive' as  
fluctuating Wilson loops (or lines) \ie representatives of the
appropriate measures (\eg $\D\Cu$).

From the previous section, in the large $N$ limit, 
\defWS\ collapses from a `gas' of Wilson loops
to just a single path integral over one Wilson loop. 
We can write this explicitly
as a path integral
over a single particle circulating in a loop:
$$S=P\int\!\!\D x\ {\rm e}^{-s[\x]+\oint\! dx.A}\quad,$$
where we have parametrized the closed curve $\Cu$ by $x^\mu(\tau)$, and
the (parametrization independent) 
path integral measure $\D x \, {\rm e}^{-s}$ is determined by the
flow equation. Note that \fflow\ indicates that this measure is
rather unusual compared to that \eg of a free particle, but 
nevertheless we see that
gauge theory in the large $N$ limit, is equivalent to a form of quantum
mechanics for a single particle!

Combining \defWS\ and \ft, yields Wilson loop expressions for the
vertices, \eg the single trace vertices are given as \alg
\eqn\npoint{ 
S_{\mu_1\cdots\mu_n}(\p_1,\cdots,\p_n)\ (2\pi)^D \delta(\sum_{i=1}^n p_i)=
(-i)^n\int\!\!\D\Cu\!\!\mathop{\oint\!\oint\!\cdots\oint}_{\displaystyle
(1,2,\cdots,n)}\!  dx_1^{\mu_1}dx_2^{\mu_2}\cdots
dx_n^{\mu_n}\ \e{-i\sum_i\x_i.\p_i}} 
(similarly $\hS$ and ${\hat\Cu}$)
where the notation $(1,2,\cdots,n)$ stands for integrating round the loop
while preserving the cyclical order, and similarly \wili\ and \wv\
give \eqnn\wnpoint 
$$\displaylines{
W_{\mu_1\cdots\mu_n,\nu_1\cdots\nu_m}(p_1,\cdots,p_n;q_1,\cdots,q_m;r,s)
\ (2\pi)^D \delta(\sum_{i=1}^n p_i+\sum_{j=1}^mq_j+r+s) =\hfill\wnpoint\cr
\hfill(-i)^{n+m}\int\!\!\!\!\int\!\!d^D\!u\,d^D\!v\int\!\!\D\Cu_{uv}
\int_u^v\!\!\!\!dx^{\mu_n}_n\!\int_u^{x_n}\!\!\!\!\!\!\!dx^{\mu_{n-1}}_{n-1}
\cdots\int_u^{x_2}\!\!\!\!\!\!\!dx^{\mu_1}_{1}
\int_v^u\!\!\!\!dy^{\nu_m}_m\!\int_v^{y_m}\!\!\!\!\!\!\!dy^{\nu_{m-1}}_{m-1}
\cdots\int_v^{y_2}\!\!\!\!\!\!\!dy^{\nu_1}_{1}\hfill\cr
\hfill \exp{-i\left(r.u+s.v+\sum_i\p_i.x_i+\sum_jq_j.y_j\right)}\quad,\cr
}$$
where the $x_i$ integration is along the curve $\Cu_{uv}$, and the
$y_j$ integration along the same curve but in the opposite direction
\cf \winexp\ or \fig\winemom{wine with momentum labels as in (618.5)}.
\midinsert
$$
\epsfxsize=0.4\hsize\epsfbox{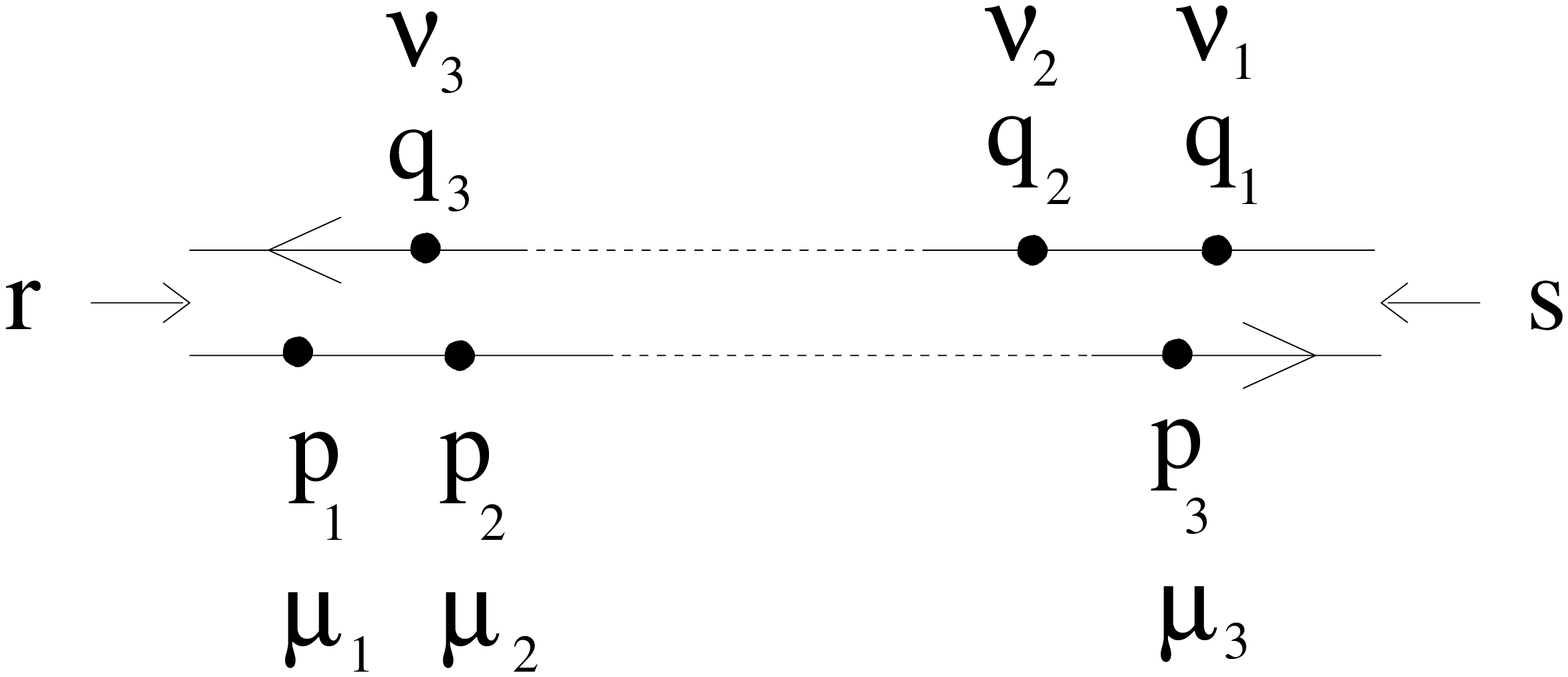}
$$
\centerline{{\bf Fig.8.} Feynman rule for the wine, with momentum labels.}
\endinsert

As we will show now in some small examples,
these Wilson loop (line) representations allow one easily to extract
many general properties of the solutions. 

\newsec{Symmetries}

The gauge invariance (trivial Ward)
identities follow by applying $\delta A_\mu
=\partial_\mu\omega-i[A_\mu,\omega]$ to \Sex\ and \wv, or more
simply by direct integration in \npoint\ and \wnpoint:
\eqn\ga{p_1^{\mu_1}S_{\mu_1\cdots\mu_n}(\p_1,\cdots,\p_n)=
S_{\mu_2\cdots\mu_n}(\p_1+\p_2,\p_3,\cdots,\p_n)
-S_{\mu_2\cdots\mu_n}(\p_2,\cdots,\p_{n-1},\p_n+\p_1).}
Note that it is clear from \Sex\ or \npoint\ that these vertices are
cyclically symmetric:
$$S_{\mu_1\cdots\mu_n}(\p_1,\cdots,\p_n)=
S_{\mu_2\cdots\mu_n\mu_1}(\p_2,\cdots,\p_n,\p_1)\quad.$$
Diagrammatically, \ga\
corresponds to a `push forward' minus a  `pull back' of the point concerned
to the end points of the relevant integration domain in \npoint,
as in \fig\fga{alg fig7 graphic gauge id }.
\midinsert
$$
\epsfxsize=.63\hsize\epsfbox{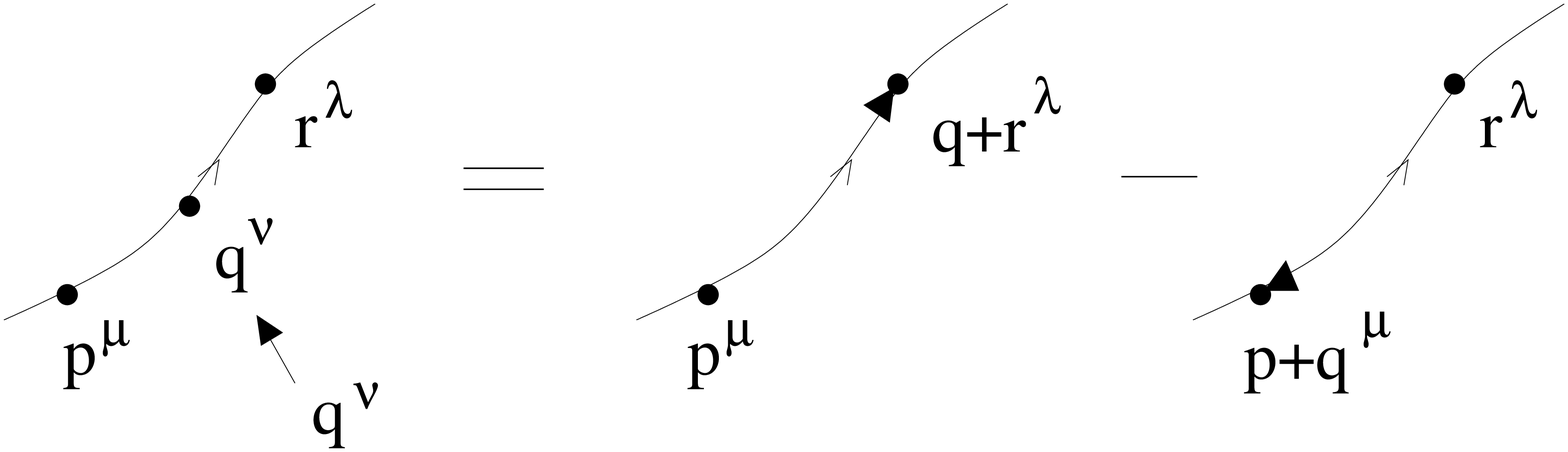}
$$
\centerline{ {\bf Fig.9.} Graphical representation of gauge invariance
identities.}
\endinsert
Of course, these same properties hold for 
$\hS_{\mu_1\cdots\mu_n}(\p_1,\cdots,\p_n)$, and separately
for each 
string in $S_{\mu_1\cdots\mu_n,\nu_1\cdots\nu_m}(p_1,\cdots,
p_n;q_1,\cdots,q_m)$, the two-trace vertices of \Sex.
(In addition
$$S_{\mu_1\cdots\mu_n,\nu_1\cdots\nu_m}(p_1,\cdots,
p_n;q_1,\cdots,q_m)=S_{\nu_1\cdots\nu_m,\mu_1\cdots\mu_n}(q_1,\cdots,
q_m;p_1,\cdots,p_n)$$
without loss of generality, from \Sex\ or \defWS.)
For the wine vertices we have the analogous identites except that 
outer momenta can reach the end of line: \eqnn\wga
$$\displaylines{p_1^{\mu_1}
W_{\mu_1\cdots\mu_n,\nu_1\cdots\nu_m}(p_1,\cdots,p_n;q_1,\cdots,q_m;r,s)=
\hfill\wga\cr
\hfill W_{\mu_2\cdots\mu_n,\nu_1\cdots\nu_m}(p_1\!+\!p_2,p_3,
\cdots,p_n;q_1,\cdots,q_m;r,s)\quad\cr
\hfill-W_{\mu_2\cdots\mu_n,\nu_1\cdots\nu_m}
(p_2,\cdots,p_n;q_1,\cdots,q_m;r\!+\!p_1,s)\quad,\cr
}$$
with similar identities for contraction with $p_n^{\mu_n}$, $q_1^{\nu_1}$
and $q_m^{\nu_m}$, as is clear from \winemom.

Charge conjugation (C) 
invariance follows from the symmetry $A_\mu\leftrightarrow-A^T_\mu$,
equivalently from reversal of the direction of all Wilson loops (lines),
thus
$$S_{\mu_1\cdots\mu_n}(\p_1,\cdots,\p_n)=
(-)^nS_{\mu_n\cdots\mu_1}(\p_n,\cdots,\p_1)\quad,$$
(similarly $\hS$) and combining C invariance with
the exchange identity,
$$W_{\mu_1\cdots\mu_n,\nu_1\cdots\nu_m}(p_1,\cdots,p_n;q_1,\cdots,q_m;r,s)
=W_{\nu_1\cdots\nu_m,\mu_1\cdots\mu_n}(q_1,\cdots,q_m;p_1,\cdots,p_n;s,r)$$
(which one easily sees by rotating \winemom\ or from 
\wc\ under $u\leftrightarrow v$), gives
$$\displaylines{
W_{\mu_1\cdots\mu_n,\nu_1\cdots\nu_m}(p_1,\cdots,p_n;q_1,\cdots,q_m;r,s)=
\hfill\cr
\hfill(-)^{n+m}\,
W_{\mu_n\cdots\mu_1,\nu_m\cdots\nu_1}(p_n,\cdots,p_1;q_m,\cdots,q_1;s,r)
\quad.}$$

Lorentz invariance implies that 
changing the sign of all the momentum arguments in the vertices
(of $S$, $\hS$ or $W$) just changes the sign of those with
an odd number of momentum arguments 
and leaves alone those with an even number. 

Finally the `coincident line' identities readily follow from \wnpoint\ by
changing the direction of the $y$ integration:
\eqn\coids{
W_{\mu_1\cdots\mu_n,\nu_1\cdots\nu_m}(p_1,\cdots,p_n;q_1,\cdots,q_m;r,s)
=(-)^m\!\!\!\!\!\sum_{interleaves}\!\!\!\!\!
W_{\la_1\cdots\la_{m+n}}(k_1,\cdots,k_{m+n};r,s)\quad,}
where we have used \compac, and the sum runs over all interleaves of
the sequences $p_1^{\mu_1},\cdots,p_n^{\mu_n}$ and 
$q_m^{\nu_m},\cdots,q_1^{\nu_1}$ \ie combined 
sequences $k_1^{\la_1},\cdots,k_{m+n}^{\la_{m+n}}$ in which the $p^\mu$s 
remain ordered with respect to each other, 
and similarly the $q^\nu$s remain in reverse order. 

Some useful examples are: \hfill\break
\centerline{$S_{\mu\nu\la}(p,q,r)$ is antisymmetric 
under exchange of any pair 
in $(p^\mu,q^\nu,r^\la)$\quad,}
\eqn\examsy{\eqalign{
W_\mu(p;q,r) &=-W_\mu(p;r,q)\quad,\qquad
p^\mu W_\mu(p;q,r) = W_q-W_r\quad,\cr
S_{\mu\nu\la\si}(p,q,r,s) &=S_{\nu\mu\si\la}(q,p,s,r)\quad,\qquad
W_{\mu\nu}(p,q;r,s)=W_{\nu\mu}(q,p;s,r)\quad,\cr
W_{;\nu_1\cdots\nu_m}(;q_1,\cdots,q_m;r,s) &=
W_{\nu_1\cdots\nu_m}(q_1,\cdots,q_m;s,r)\quad,\cr
W_{\mu,\nu}(p;q;r,s) &= -W_{\mu\nu}(p,q;r,s)-W_{\nu\mu}(q,p;r,s)\quad.\cr}}

\newsec{Perturbative expansion}

Although we will be especially interested in four dimensions, 
we keep $D\ne4$ here since it will be helpful to access this
via the limit $D\to4$ \alg.
It will also be helpful to write \RG\ as
\eqn\ERG{\Lambda{\partial\over\partial\Lambda}S=
-a_0[S,g^2S-2\hS]+a_1[g^2S-2\hS]\quad,}
where we have used \Sig\ and written the classical
term as the bilinear functional $-a_0$ 
and the quantum term as the linear functional 
$a_1$.\footnote{${}^\dagger$}{The subscripts indicate the order of $\hbar$,
\ie the number of loops involved.} 
At the classical level, we
see from \defg\ and \ERG, that $S\sim 1/g^2$. Substituting
this in the quantum term of \ERG, we see that
the one loop correction $\sim g^0$, and so on.
Thus by iteration, $S$ has the weak coupling expansion$^\dagger$
\eqn\Sloope{S={1\over g^2} S_0+S_1+g^2 S_2 +\cdots\quad.}
This is of course the form expected by the usual graphical considerations
[after performing $A_\mu\mapsto A_\mu/g$ as above \poluii].
Substituting this expansion in \ERG\ and recalling that $g$ will run at
the quantum level, we see that
the $\beta$ function must also take the standard form$^\dagger$
\eqn\betafn{\beta:=\Lambda{\partial g\over\partial\Lambda}=\beta_1g^3+\beta_2g^5+\cdots\quad.}
We will see below how the $\beta_i$ are determined. 
From  \Sloope\ and \betafn, we obtain the loopwise expansion
of \ERG:
\eqnn\ergcl\eqnn\ergone\eqnn\ergtwo
$$\eqalignno{\Lambda{\partial\over\partial\Lambda}S_0 &=-a_0[S_0,S_0-2\hS]
&\ergcl\cr
\Lambda{\partial\over\partial\Lambda}S_1&=2\beta_1S_0-2a_0[S_0-\hS,S_1]
+a_1[S_0-2\hS] &\ergone\cr
\Lambda{\partial\over\partial\Lambda}S_2&=2\beta_2S_0-2a_0[S_0-\hS,S_2]
-a_0[S_1,S_1]+a_1[S_1]\quad, &\ergtwo\cr}$$
\etc We mention in passing
that by considering linear perturbations $S\mapsto S+
\epsilon {\cal O}$, we easily obtain the flow equations
for the integrated operator ${\cal O}$, and its
weak coupling expansion. Thus from \ERG,
\eqn\OP{\Lambda{\partial\over\partial\Lambda}\O=
-2a_0[g^2S-\hS,\O]+g^2a_1[\O]\quad.}
From \Sloope, we expand $\O={1\over g^2}\O_0+\O_1+g^2\O_2+\cdots$ giving,
either from \betafn, \Sloope\ and \ERG, or directly from \ergcl\ -- \ergtwo:
\eqnn\opcl\eqnn\opone\eqnn\optwo
$$\eqalignno{\Lambda{\partial\over\partial\Lambda}\O_0 &=-2a_0[S_0-\hS,\O_0]
&\opcl\cr
\Lambda{\partial\over\partial\Lambda}\O_1&=2\beta_1\O_0-2a_0[S_0-\hS,\O_1]
-2a_0[S_1,\O_0]
+a_1[\O_0] &\opone\cr
\Lambda{\partial\over\partial\Lambda}\O_2&=2\beta_2\O_0-2a_0[S_0-\hS,\O_2]
-2a_0[S_2,\O_0]
-2a_0[S_1,\O_1]+a_1[\O_1]\quad, &\optwo\cr}$$
\etc

\subsec{Feynman diagrammatics and the classical vertices}

Our diagrams, as in \fflow, play one final r\^ole: expanded as in \winexp,
they yield Feynman diagrams which may be translated directly into the
corresponding equations for the vertices. The translation 
rule is just that we must place the points and their associated 
momenta in all places on composite Wilson loops, 
whilst preserving the cyclic order. We then read off the
appropriate wine and action vertices (applying momentum conservation --and
including momentum integrals if appropriate), contract 
the Lorentz indices of
points joined by wines and, from \RG, multiply the whole by $1/2\Lambda^2$.
In the following we compute 
in this way the first few classical vertices\foot{For convenience 
we move the $\hbar$ counting index to a superscript}
$S^0_{\mu_1\cdots\mu_n}$.

The diagrams for the
classical two-point vertex $S^0_{\mu\nu}(p)$ follow from \ergcl\ and 
the top two lines of \fflow:
$$
\Lambda{\partial\over\partial\Lambda}\,
\mathop{\vcenter{\epsfxsize=0.085\hsize\epsfbox{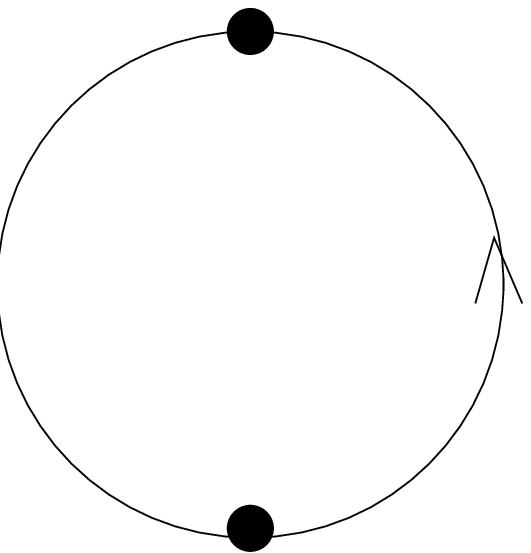}}^{\phantom{p}}
}^{\displaystyle
-p^\nu}_{\displaystyle p^\mu}=\ 2\!\!\!\!\!\!
\mathop{\vcenter{\epsfxsize=0.13\hsize\epsfbox{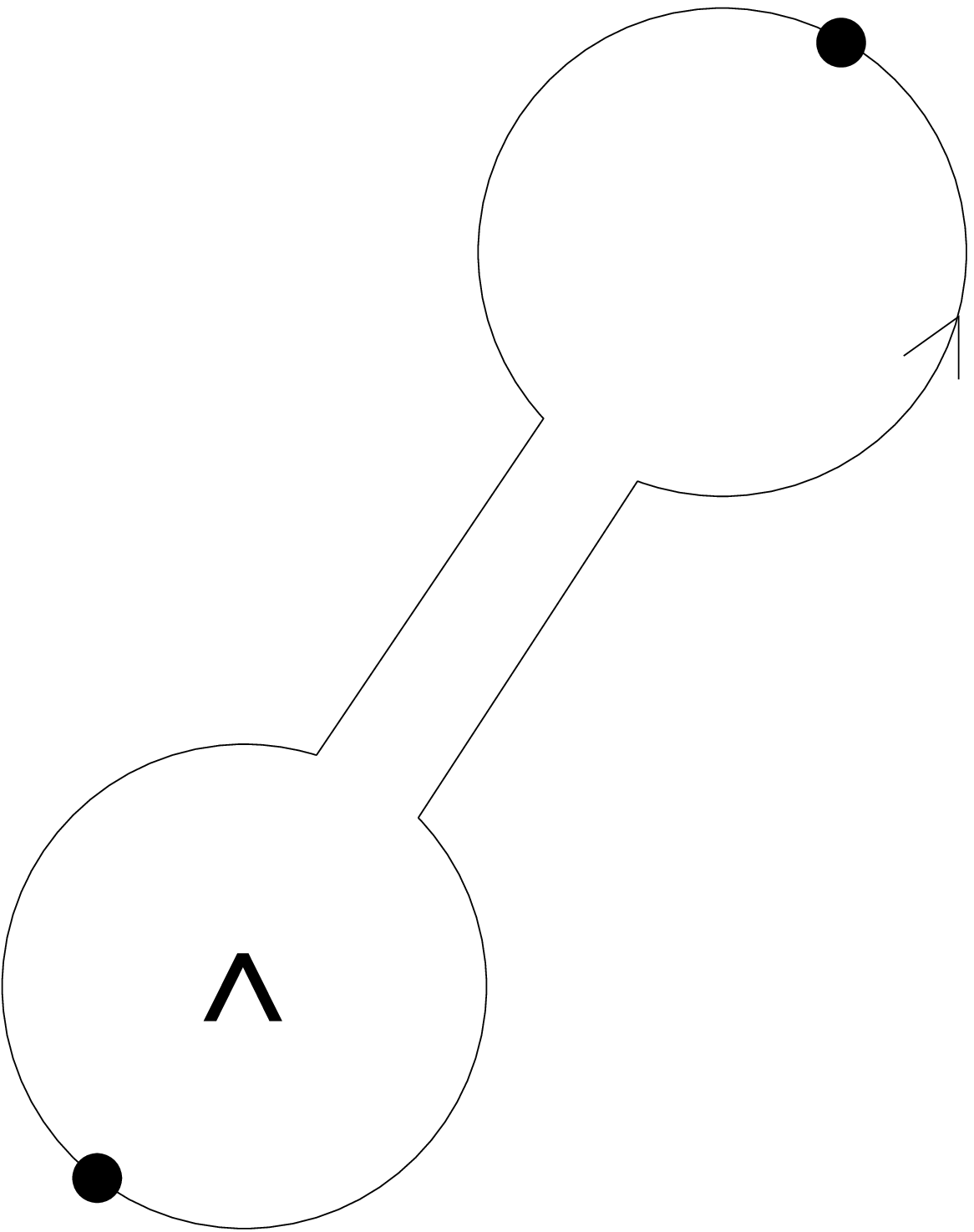}}}_{
\!\!\!\!\!\!\!\!\!\!\!\!\!\!\!\!\displaystyle p^\mu}
^{\displaystyle\;\;\;\;\;\;\;\;\;\;\;\;\;\;\;\;\;\;\;\;-p^\nu}
-\!\!\!\!\!\!\mathop{\vcenter{\epsfxsize=0.13\hsize\epsfbox{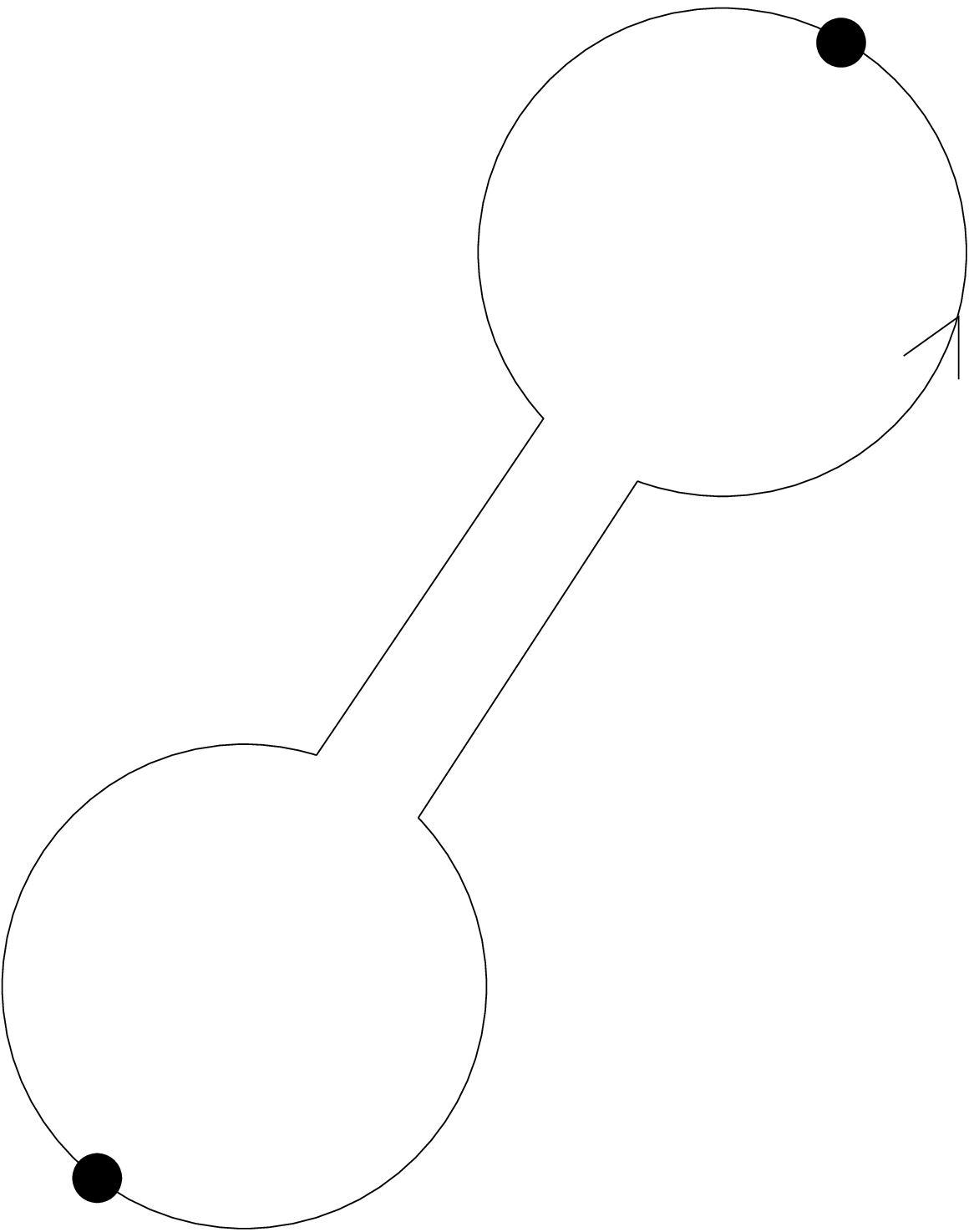}}}_{
\!\!\!\!\!\!\!\!\!\!\!\!\!\!\!\!\displaystyle p^\mu}
^{\displaystyle\;\;\;\;\;\;\;\;\;\;\;\;\;\;\;\;\;\;\;\;-p^\nu}
+(p^\mu\leftrightarrow -p^\nu)$$
\centerline{ {\bf Fig.10.} Feynman diagrams for the two-point vertex.}
\nfig\ftwo{fig.12 of Alg two-point fl}
Here, the empty circle corresponds to $S_0$, not $S$ as in \fflow,
and we have noted that since actions' one-point 
vertices vanish (by $\tr A_\mu=0$ but see also \alg),
we must have at least one blob per lobe. Thus
\eqn\dtwop{\Lambda{\partial\over\partial\Lambda}S^0_{\mu\nu}(p)=
-{1\over2\Lambda^2}c'_p
\left[S^0_{\mu\lambda}(p)-2\hS_{\mu\lambda}(p)\right]
S^0_{\lambda\nu}(p)+(p_\mu\leftrightarrow -p_\nu)\quad,}
where similarly to \hSex, $c'_p\equiv c'(p^2/\Lambda^2)$.
By gauge invariance and dimensions,
\eqn\partwo{S^0_{\mu\nu}(p)=2\Delta_{\mu\nu}(p)/f(p^2/\Lambda^2)\quad.}
From \Sloope, we require $f(0)=1$ so
as to be consistent with \defg\ in the $g\to0$ limit. 
Substituting \hSex, we readily find the unique solution to be $f=c$,
and thus
\eqn\twop{S^0_{\mu\nu}(p)=\hS_{\mu\nu}(p)\quad.} 

Similarly, from \ergcl\ and the top two lines of \fflow, we obtain the
following diagrams for the three-point vertex:
$$\eqalign{
\Lambda{\partial\over\partial\Lambda}\
\vcenter{\epsfxsize=0.085\hsize\epsfbox{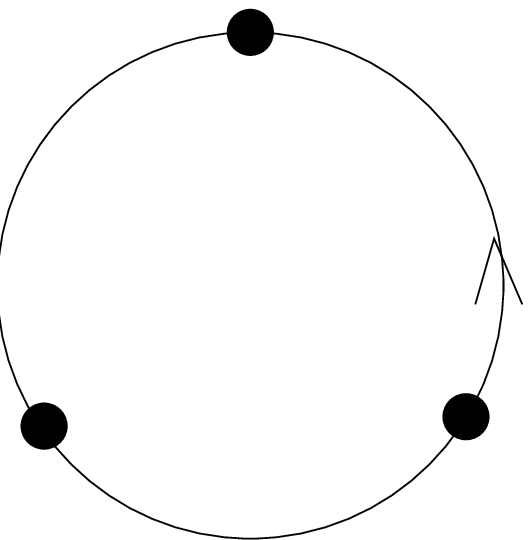}}\,
=\,\, &2\vcenter{\epsfxsize=0.13\hsize\epsfbox{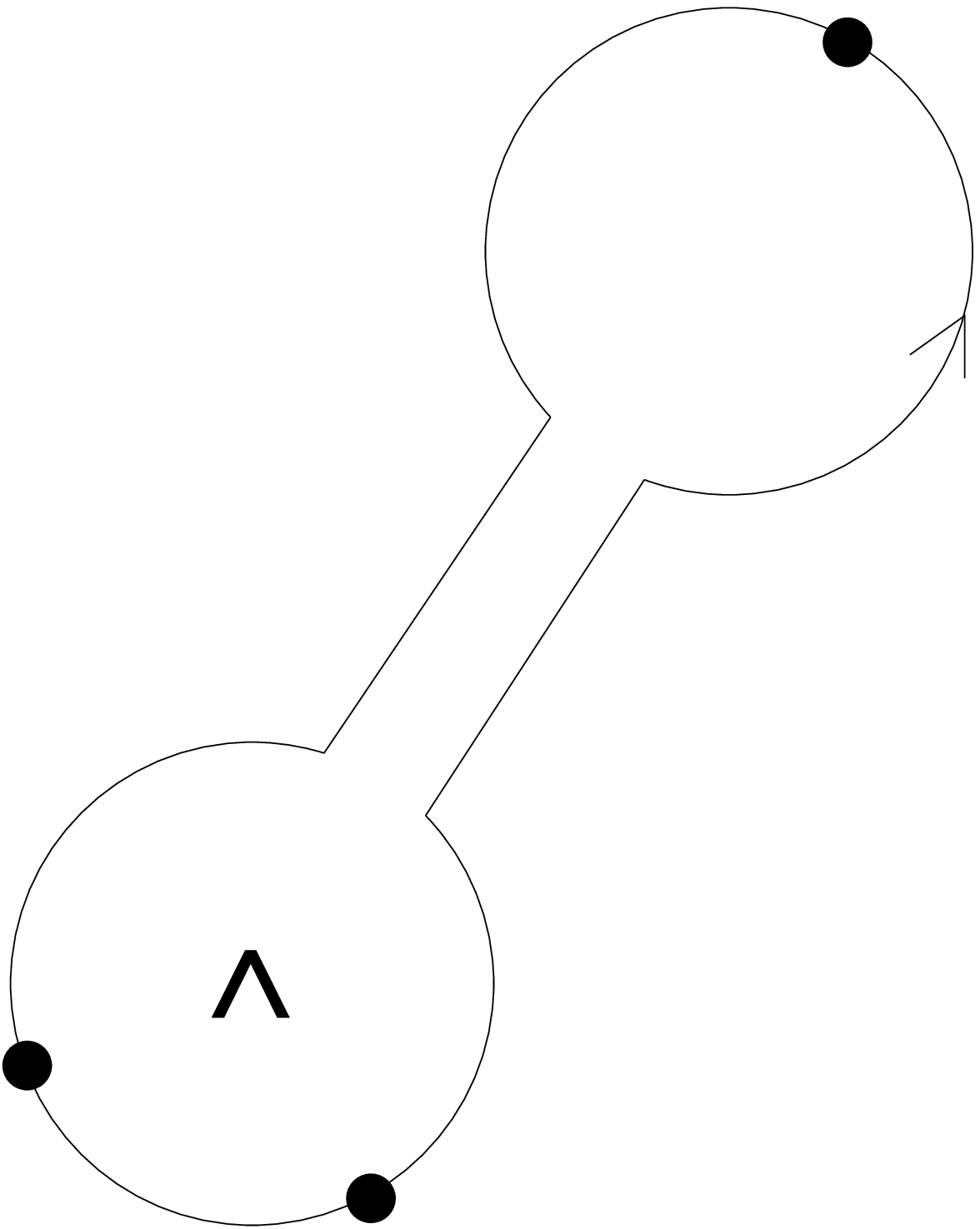}}
+2\vcenter{\epsfxsize=0.13\hsize\epsfbox{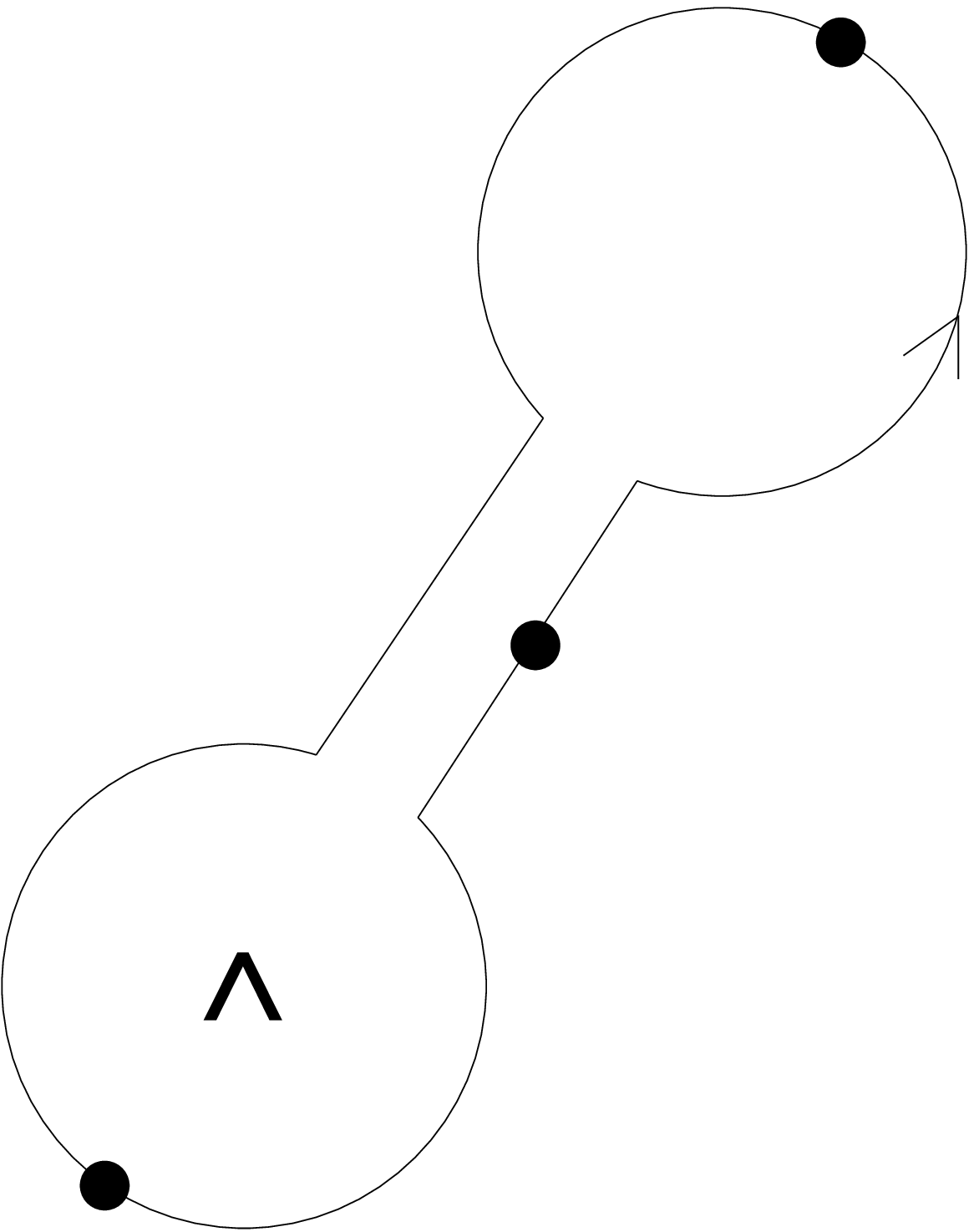}}\cr
+2\vcenter{\epsfxsize=0.13\hsize\epsfbox{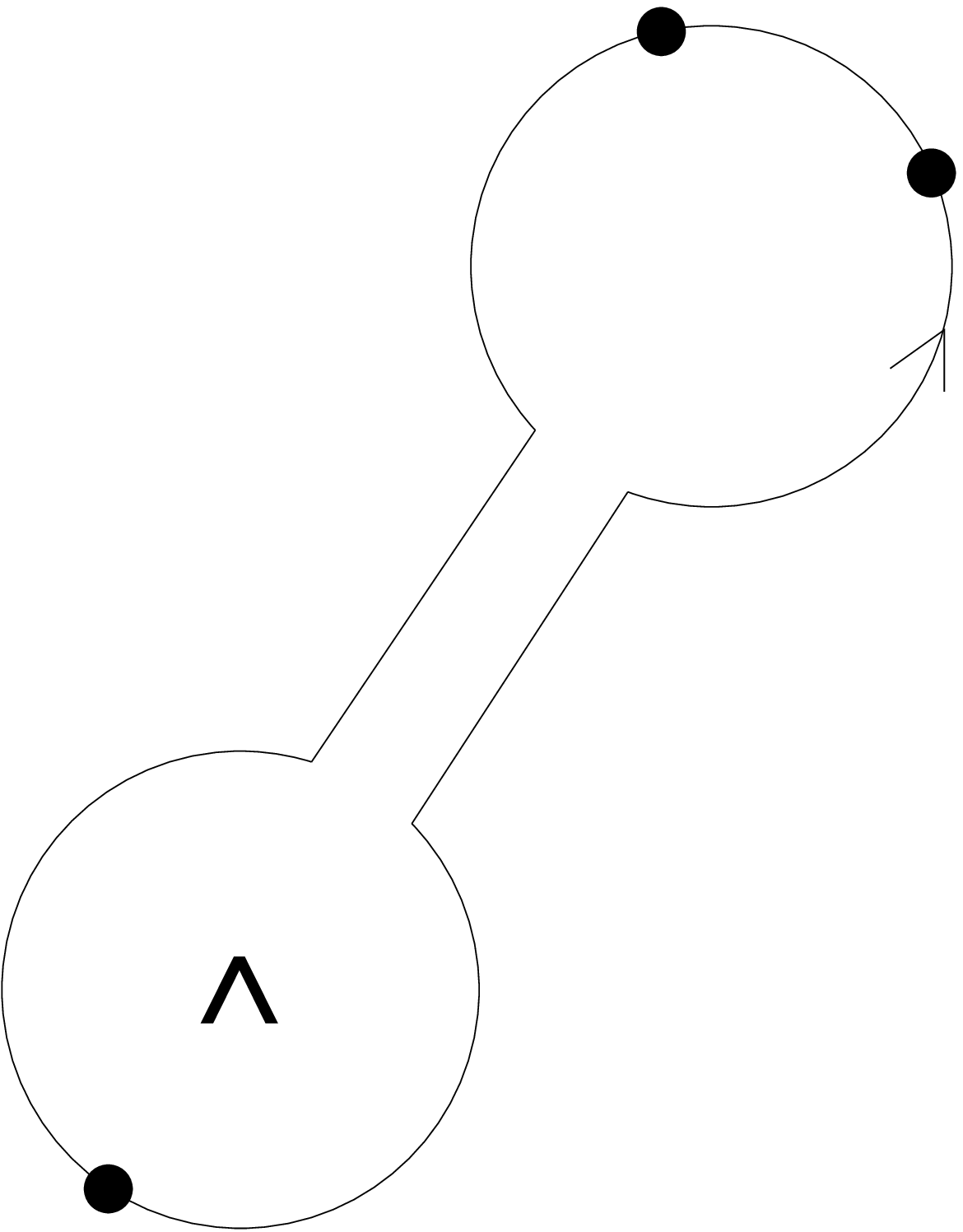}}
-2\vcenter{\epsfxsize=0.13\hsize\epsfbox{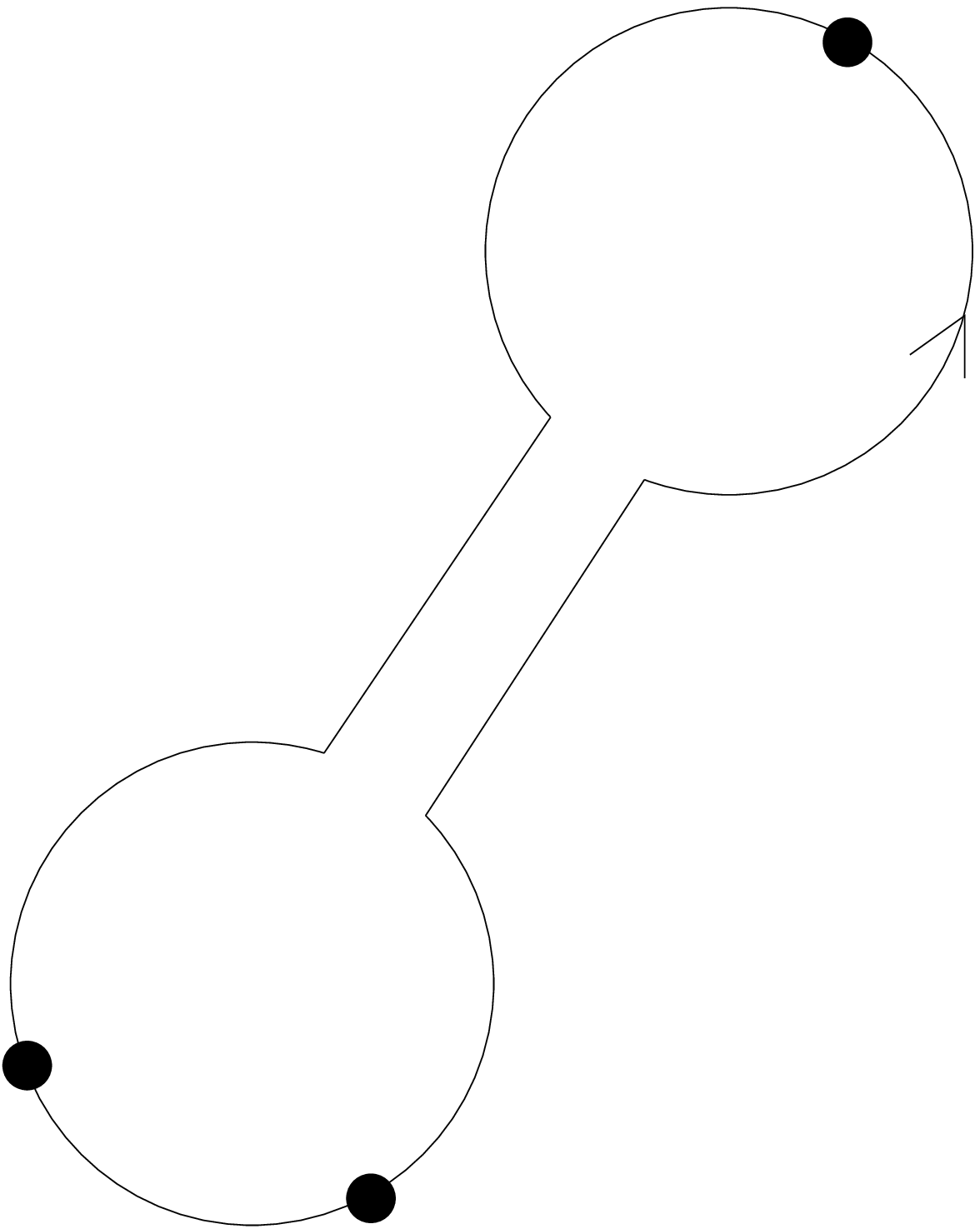}}
&+2\vcenter{\epsfxsize=0.13\hsize\epsfbox{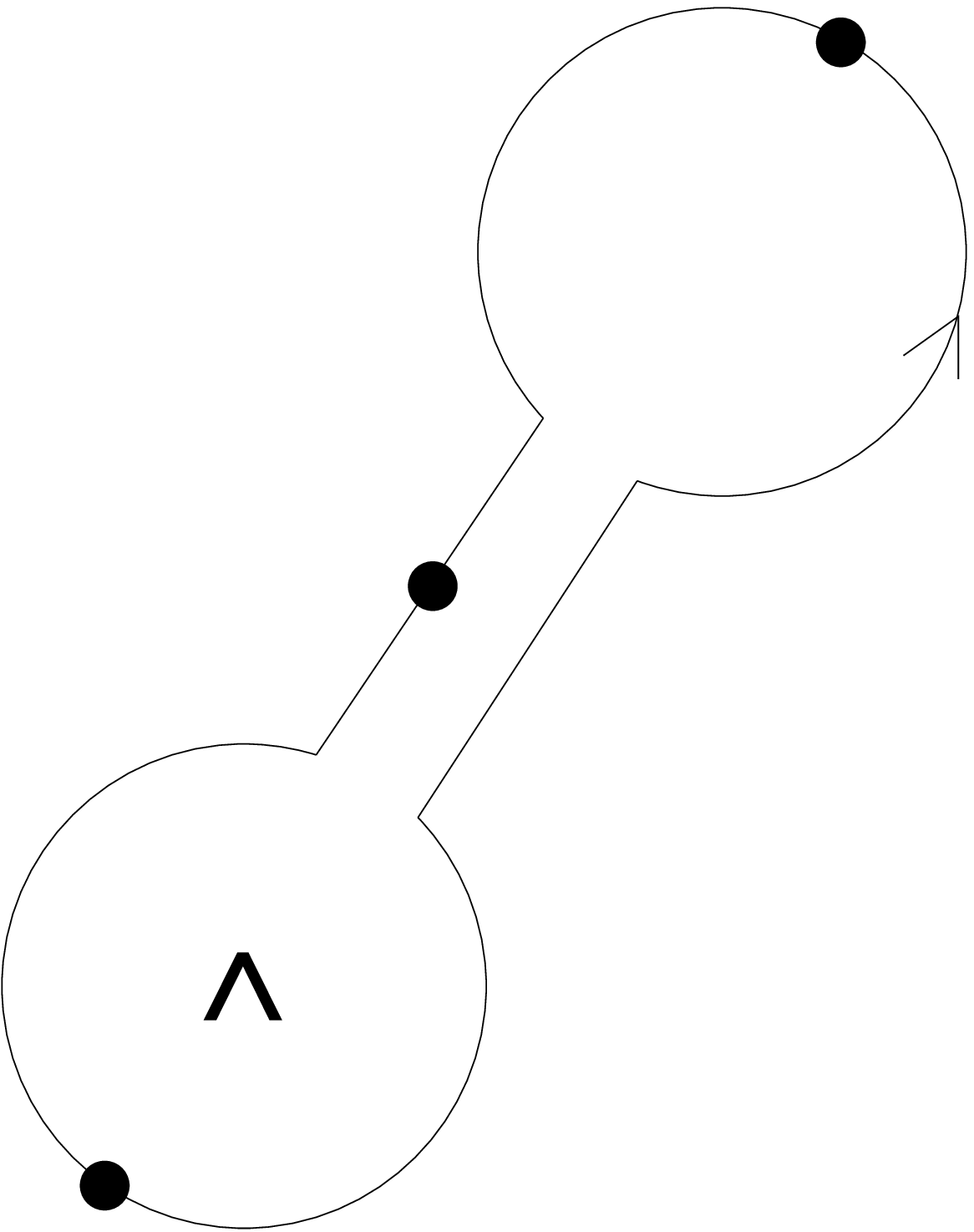}}
-2\vcenter{\epsfxsize=0.13\hsize\epsfbox{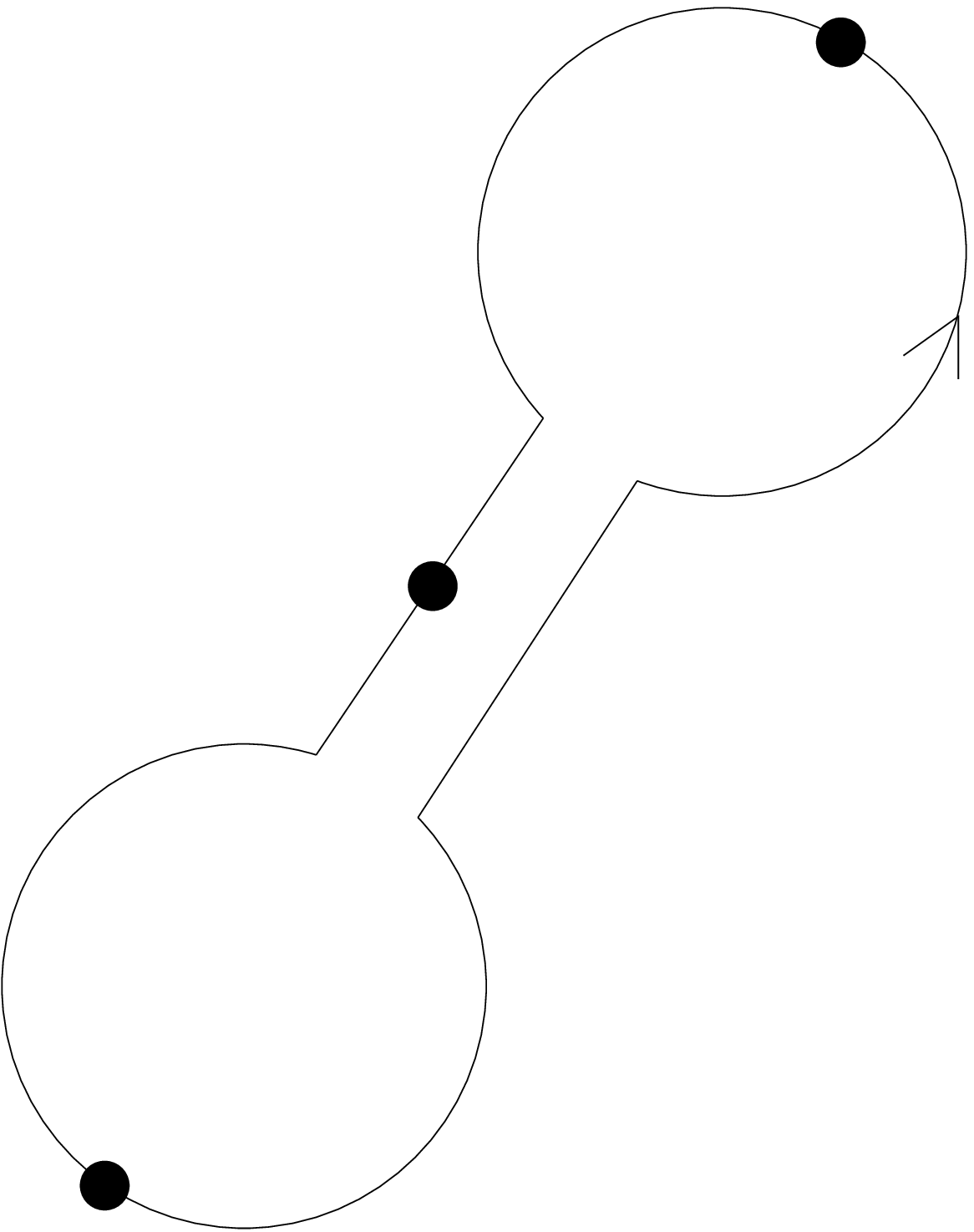}}\cr
}
$$
\centerline{ 
{\bf Fig.11.} Feynman diagrams for the three-point vertex.}
\nfig\fthree{Fig13 Alg three-point fl}
Using \twop, the entire second line vanishes and we are left on the RHS
only with terms that are already determined. We can thus immediately 
integrate to get
\eqn\threep{\eqalign{
S^0_{\mu\nu\lambda}(\p,\q,\r)= &-\int_\Lambda^\infty\!\!{d\Lambda_1\over
\Lambda_1^3}\left\{c'_r\hS_{\mu\nu\alpha}(\p,\q,\r)\hS_{\alpha\lambda}(r)
+c'_\nu(\q;\p,\r)\hS_{\mu\alpha}(p)\hS_{\alpha\lambda}(r)
\right\}\cr
&+2(r_\nu\delta_{\mu\lambda}-r_\mu\delta_{\nu\lambda})\qquad\qquad\qquad
+{\rm cycles}\quad.}}
Here it should be understood that in the curly brackets we replace $\Lambda$
with $\Lambda_1$, and to the whole expression we add the two cyclic
permutations of $(p_\mu,q_\nu,r_\lambda)$. In principle the top limit would be $\Lambda_0$, the
bare cutoff scale, but since the integrands have Taylor expansions
in momenta$/\Lambda_1$, 
the continuum limit $\Lambda_0\to\infty$ trivially
exists (as expected at the classical level). 
The integration constant is a term independent of $\Lambda$, which by
dimensions and locality (`quasi' or `ultra', \cf sec. 4) 
must be linear in the momenta.
By gauge invariance it has to be the unique covariantization
of $2\Delta_{\mu\nu}$ (see below), \ie the
three-point vertex in ${1\over2}\int\!\!d^D\!x\, F_{\mu\nu}^2$. 
In fact, this conclusion follows directly by comparing the 
$\Lambda\to\infty$ limit of \deforg\ with \threep.

Using \ga\ and \wga, one may readily check the gauge invariance of 
\threep: contracting the RHS of \threep\ with $p^\mu$, we obtain 
\eqn\gcheck{-\int_\Lambda^\infty\!\!{d\Lambda_1\over
\Lambda_1^3}\left\{ c'_r\hS_{\nu\alpha}(r)\hS_{\alpha\lambda}(r)
-c'_q\hS_{\lambda\alpha}(q)\hS_{\alpha\nu}(q)\right\}
+2\Delta_{\nu\lambda}(r)-2\Delta_{\nu\lambda}(q)\quad,}
after cancellation of some `corner'
terms containing $\hS_{\nu\alpha}(q)\hS_{\alpha\lambda}
(r)$ and $c'_r$ or $c'_q$.
From \dtwop\ and \twop, the integrand is a total derivative
integrating to boundary terms 
$\left[ \hS_{\nu\lambda}(r)-\hS_{\nu\lambda}(q)\right]^\infty_\Lambda$.
The top limit cancels the $\Delta$ terms in \gcheck\ if and only if
the integration constant in \threep\ is a gauge covariantization 
of $2\Delta_{\mu\nu}(p)$, while the bottom limit gives
$S^0_{\nu\lambda}(r)-S^0_{\nu\lambda}(q)$ evaluated at $\Lambda$,
as required. In fact, this check is  clear essentially diagrammatically, 
using \fga, \fthree, and \ftwo\ as follows. 
On the RHS of \fthree, we sum the results of pushing forward and
pulling back each point in turn (\ie summing over the cyclic
orderings of the momentum $p^\mu$).
The cancelling terms mentioned above,  
correspond to terms that meet at a corner where the wine joins $\hS$ --
push-forwards to the end of a
wine cancelling pull-backs to this point in $\hS$, and \vv\ --
while the remaining terms represent
the integrand in \gcheck. These appear as the two-point diagrams of \ftwo,
with each point in turn
replaced by the `pushforward' and `pullback' arrows of \fga.

The simplification that the flow depends only on already
known terms persists to all higher point 
classical vertices (and therefore also to all
orders in $\hbar$, \aka $g$):
Since we need at least one blob per lobe, the flow of
a classical $n$-point vertex cannot depend on a higher-point vertex,
and thus these equations are closed.
Furthermore, when one lobe corresponds to an $n$-point $S^0$-vertex, the
other corresponds to the two-point vertex \twop. 
Since for $n>2$, there are two such terms in $a_0[S^0,S^0]$
but only one in $a_0[S^0,\hS]$, these contributions cancel in \ergcl.
Thus the flow of $S^0_{\mu_1\cdots\mu_n}$ for $n>2$,
depends only on lower-point
vertices or $\hS$, and may be immediately integrated.
$$\eqalign{
\Lambda{\partial\over\partial\Lambda}\
\vcenter{\epsfxsize=0.085\hsize\epsfbox{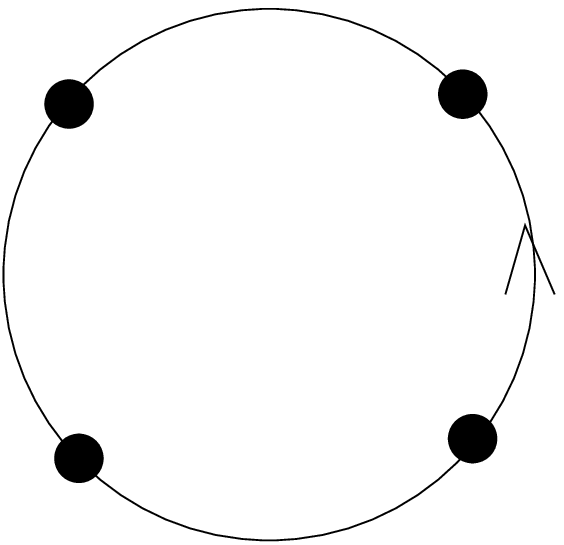}}\,
=\, -\vcenter{\epsfxsize=0.13\hsize\epsfbox{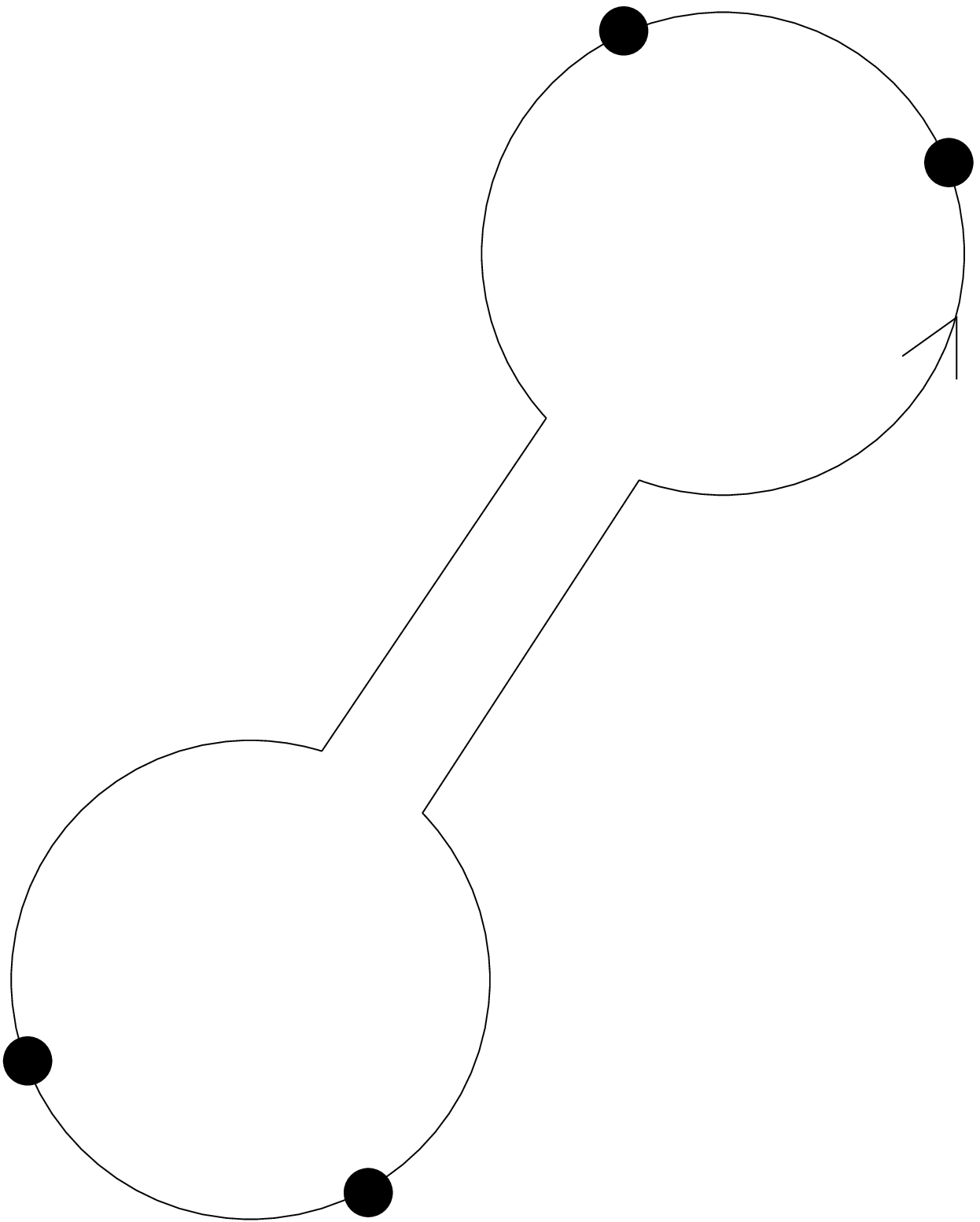}}
&+2\vcenter{\epsfxsize=0.13\hsize\epsfbox{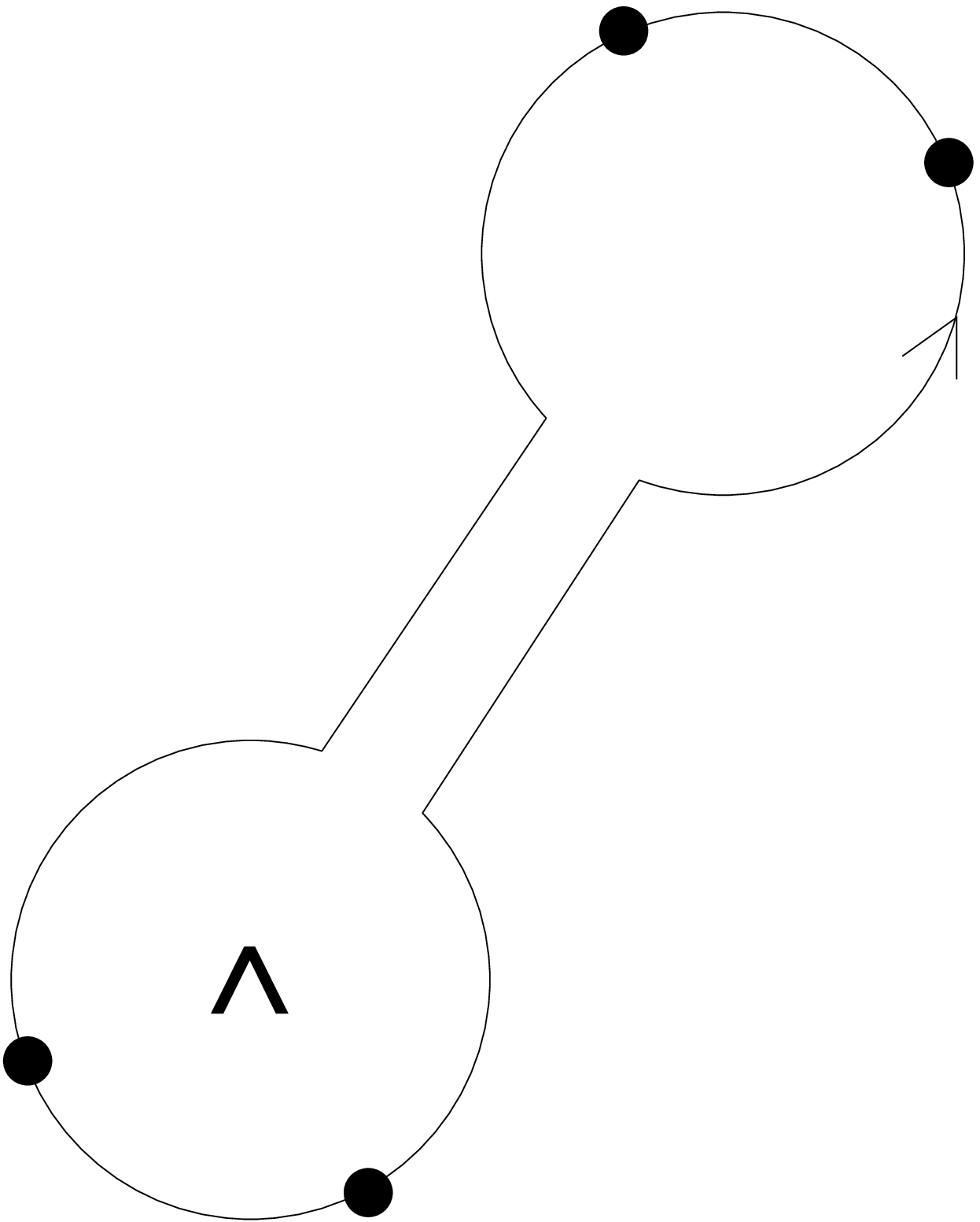}}
+2\vcenter{\epsfxsize=0.13\hsize\epsfbox{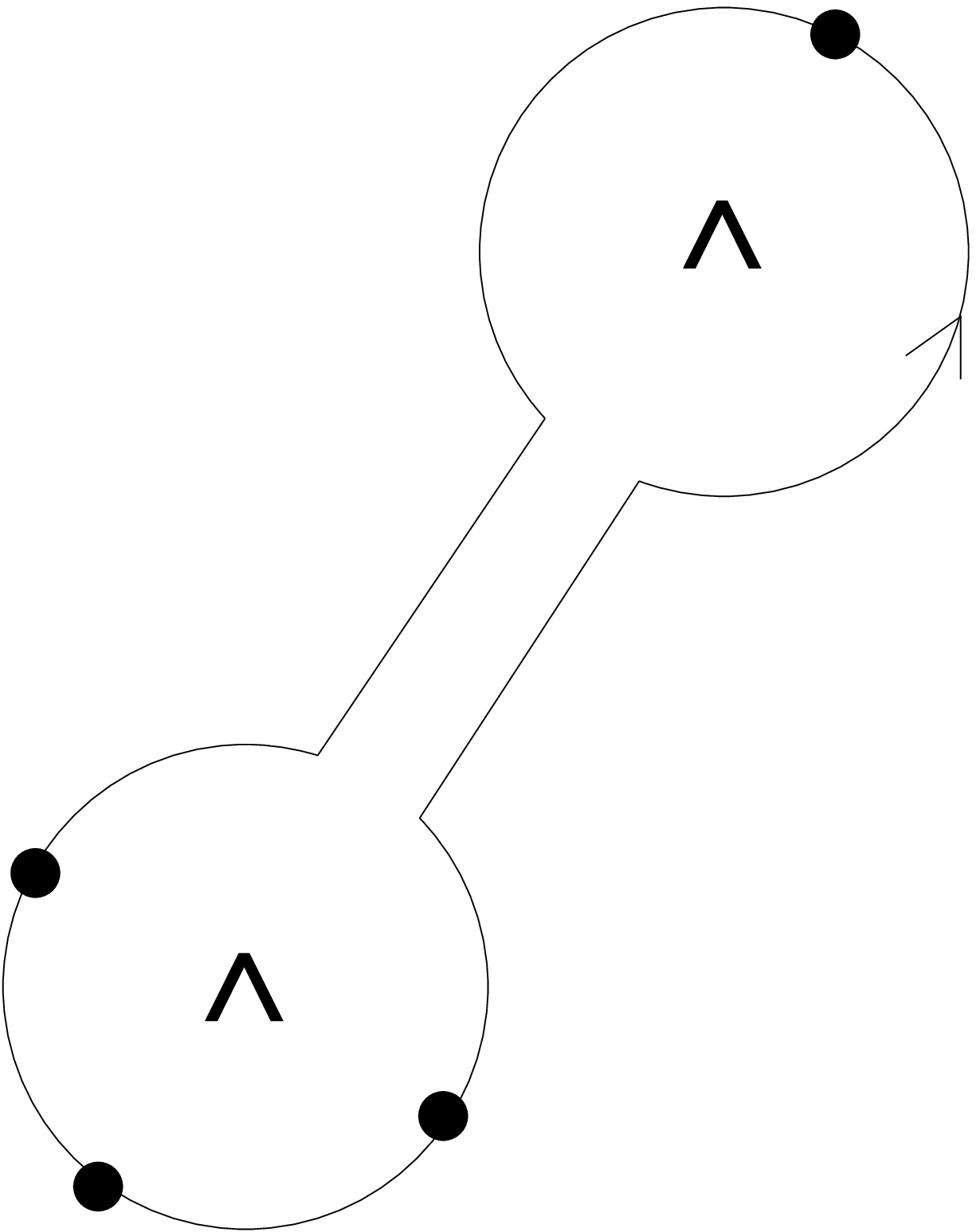}}
+2\vcenter{\epsfxsize=0.13\hsize\epsfbox{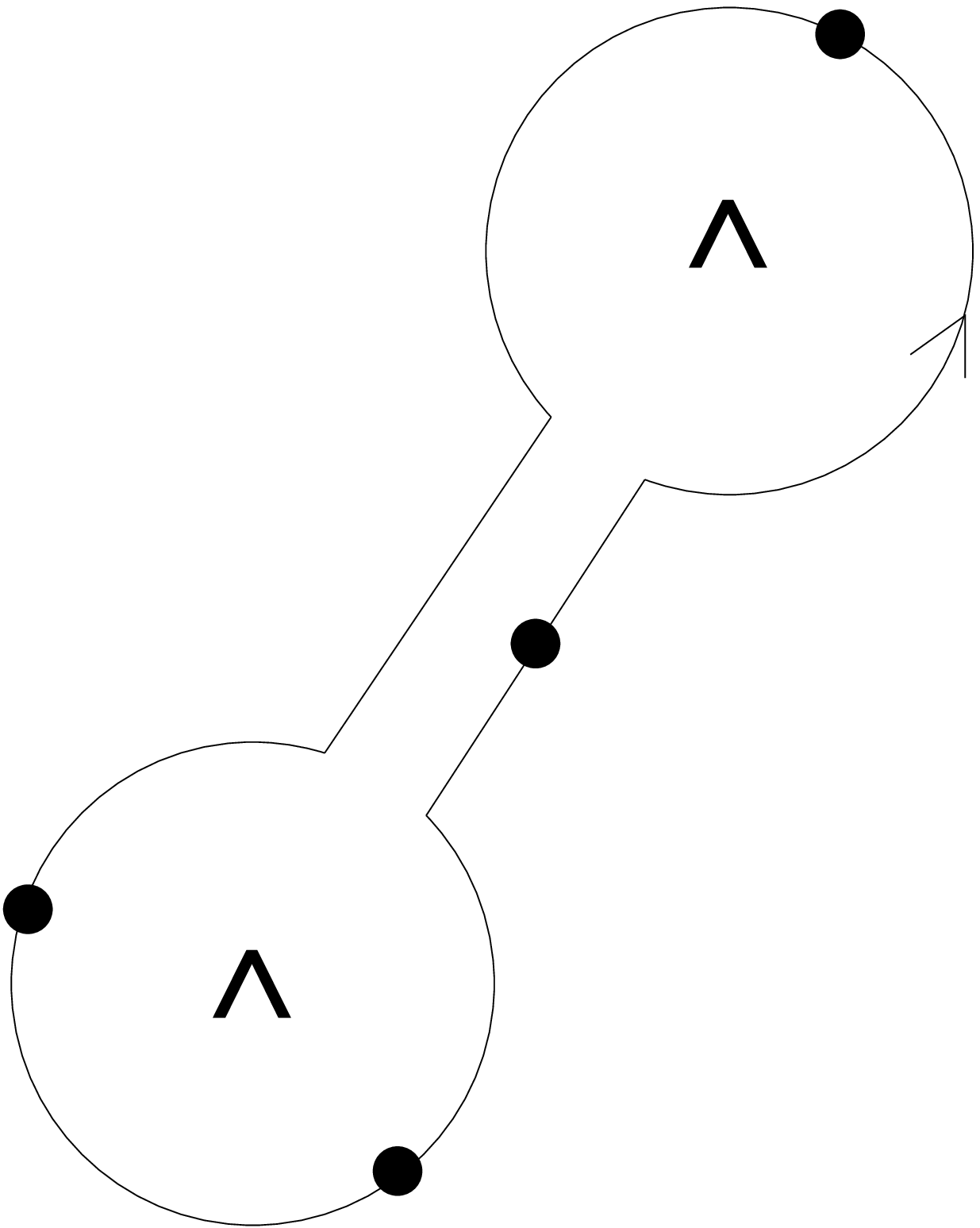}}\cr
&+2\vcenter{\epsfxsize=0.13\hsize\epsfbox{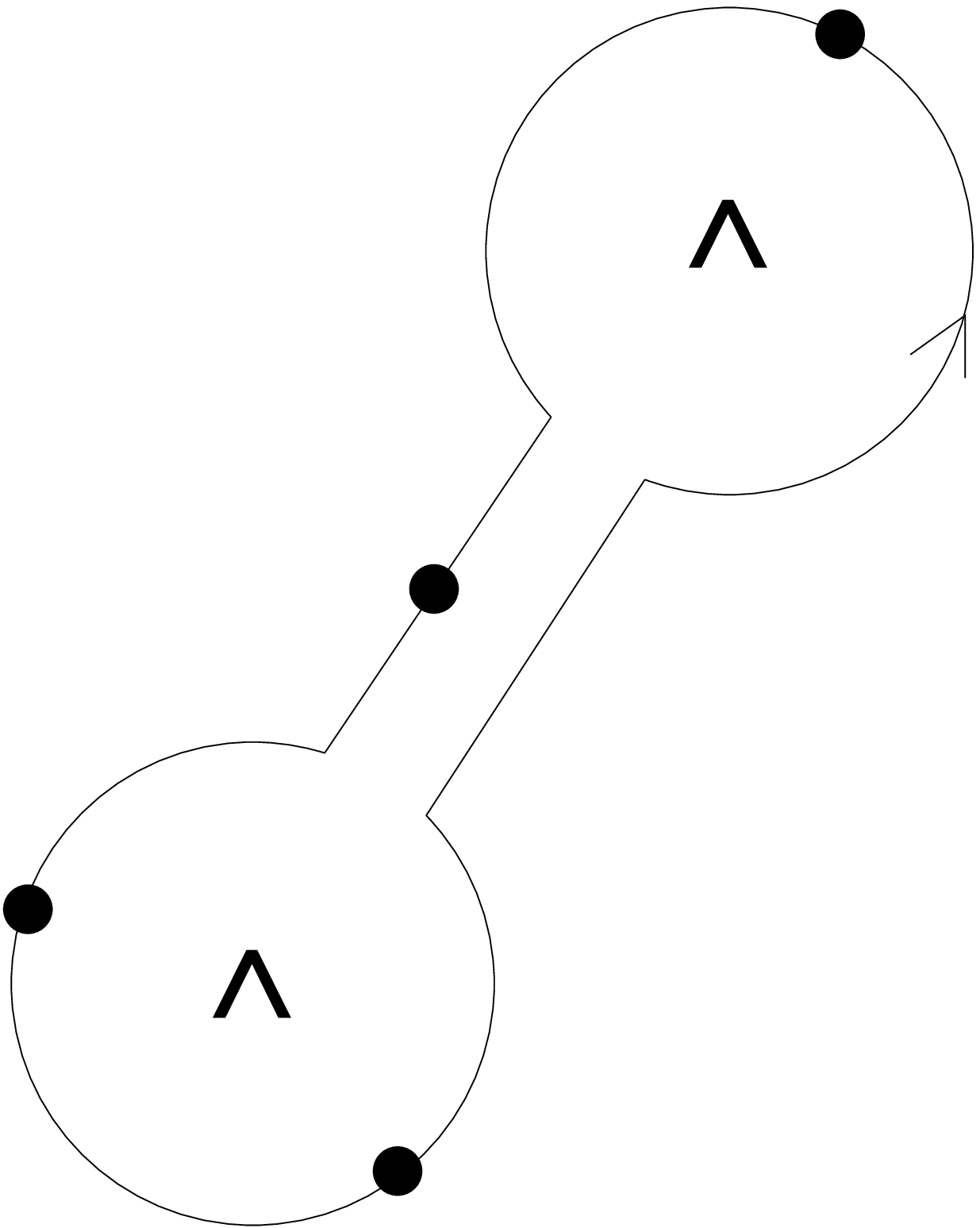}}
+2\vcenter{\epsfxsize=0.13\hsize\epsfbox{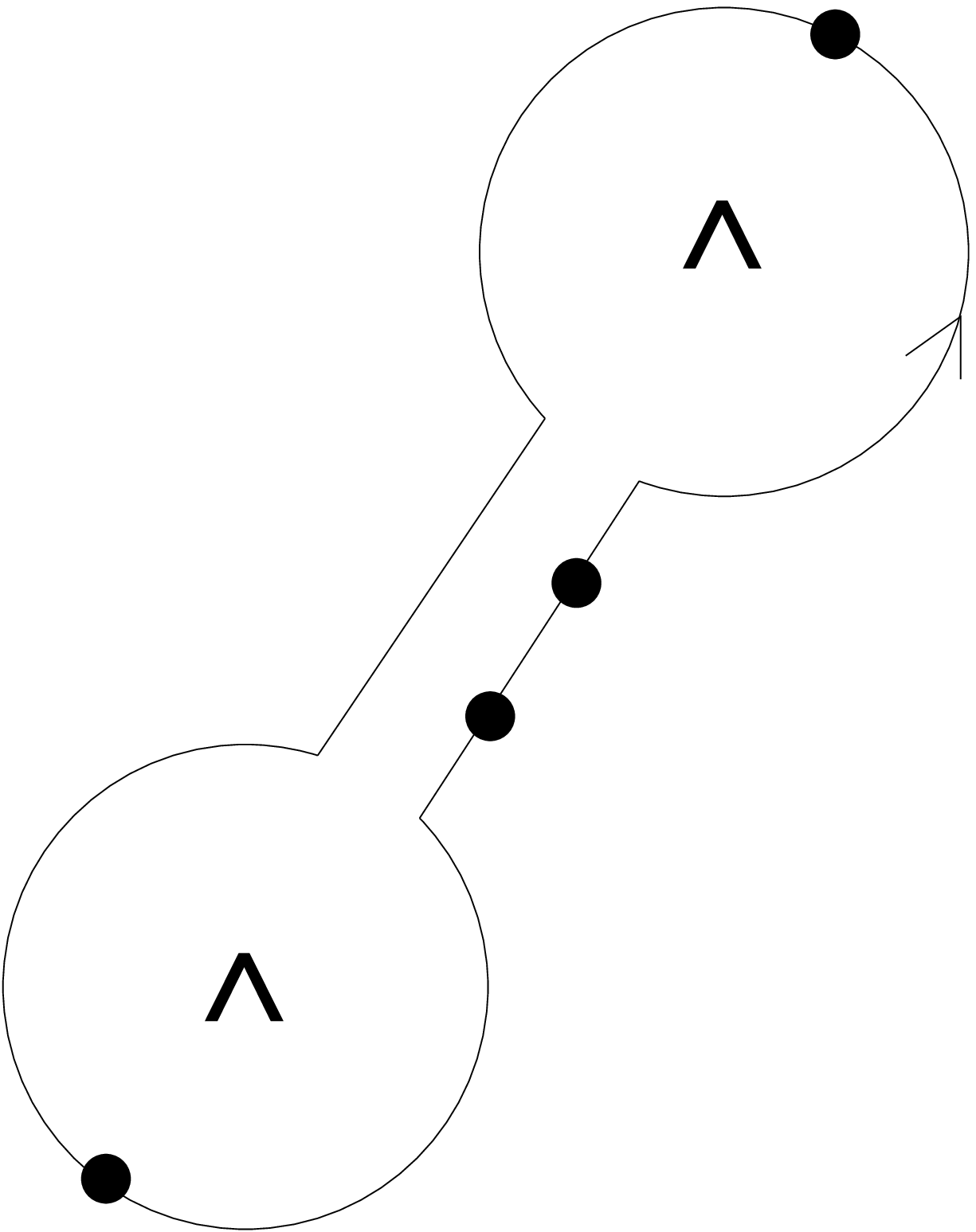}}
+\vcenter{\epsfxsize=0.13\hsize\epsfbox{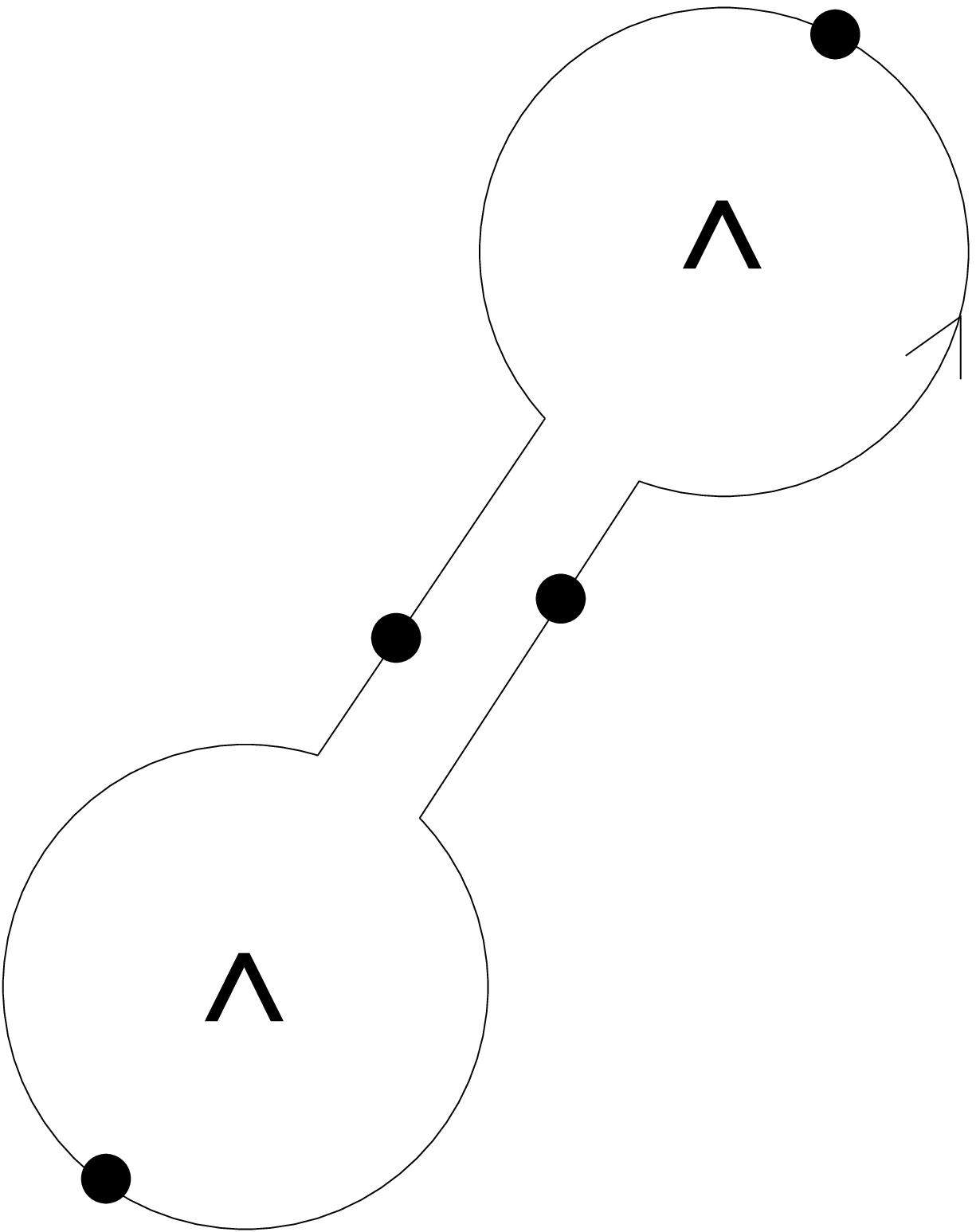}}
}
$$
\centerline{ 
{\bf Fig.12.} Feynman diagrams for the four-point vertex.}
In a similar way, we compute the classical gauge invariant continuum
four-point vertex. After simplifying by \twop,
the diagrams are those in \fig\ffour{see
p 385 and times by -. CARE to have two $S^0$ three-points first or
alter later ref to ffour!!}, thus 
\eqn\fourp{\eqalign{S^0_{\mu\nu\la\sigma}(p, &q,r,s) =
-\int_\Lambda^\infty\!\!{d\Lambda_1\over
\Lambda_1^3}\Big\{
c'_{p+q}\left(
\hS_{\mu\nu\alpha}(p,q,r\!+\!s)
-\half S^0_{\mu\nu\alpha}(p,q,r\!+\!s)
\right)S^0_{\alpha\lambda\sigma}(p\!+\!q,r,s)\cr
&+\hS_{\sigma\alpha}(s)\hS_{\mu\nu\alpha}(p,q,r\!+\!s)c'_\lambda(r;p\!+\!q,s)
+\hS_{\lambda\alpha}(r)\hS_{\mu\nu\alpha}(p,q,r\!+\!s)c'_\sigma(s;r,p\!+\!q)\cr
&+\hS_{\mu\alpha}(p)\hS_{\alpha\sigma}(s)c'_{\nu\la}(q,r;p,s)
+\half\hS_{\mu\alpha}(p)\hS_{\alpha\lambda}(r)c'_{\nu,\sigma}(q;s;p,r)\cr
&+c'_s\hS_{\sigma\alpha}(s)\hS_{\mu\nu\la\alpha}(p,q,r,s)
+ {\rm cycles} \Big\}
+2\delta_{\mu\sigma}\delta_{\nu\lambda}
-4\delta_{\mu\lambda}\delta_{\nu\sigma}+2\delta_{\mu\nu}\delta_{\lambda\sigma}
\quad,}}
where `cycles' stands for the three cyclic permutations of $(p_\mu,q_\nu,
r_\la,s_\si)$. We see that, as predicted in sect.5,
\twop, \threep\ and \fourp\ are smooth in momenta.

Whilst \fourp\
may look complicated, the expression is actually
completely determined from the top
term by gauge invariance and cyclicity:
Indeed, the top term is
sufficient to identify the two composite Wilson loops in \ergcl, and
from the graphical arguments below \gcheck, gauge invariance
then holds only if all the other terms appear 
with their correct coefficients, since these
correspond to all cyclic order preserving positions of the points 
on the underlying composite Wilson loop, and only then do all 
the `corner' cancellations correctly take place. 
Similarly all the other (\eg higher-loop and/or higher-point) $S$-vertices 
follow from just one or two of their individual terms.
In fact, since \seed\ is also a composite Wilson loop \defWhS\ \alg, 
these same comments apply to the $\hS$-vertices \hSex.

At finite $N$, classical vertices with more than one trace do exist. However,
simply by $\tr A_\mu=0$, there can be no such vertices with three or less
points. The four-point vertex $S^0_{\mu\nu,\la\si}(p,q;r,s)$
actually also vanishes. Its flow is shown diagrammatically in fig. 13,
where once again we have simplified with \twop.
$$\Lambda{\partial\over\partial\Lambda}\
\vcenter{\epsfxsize=0.1\hsize\epsfbox{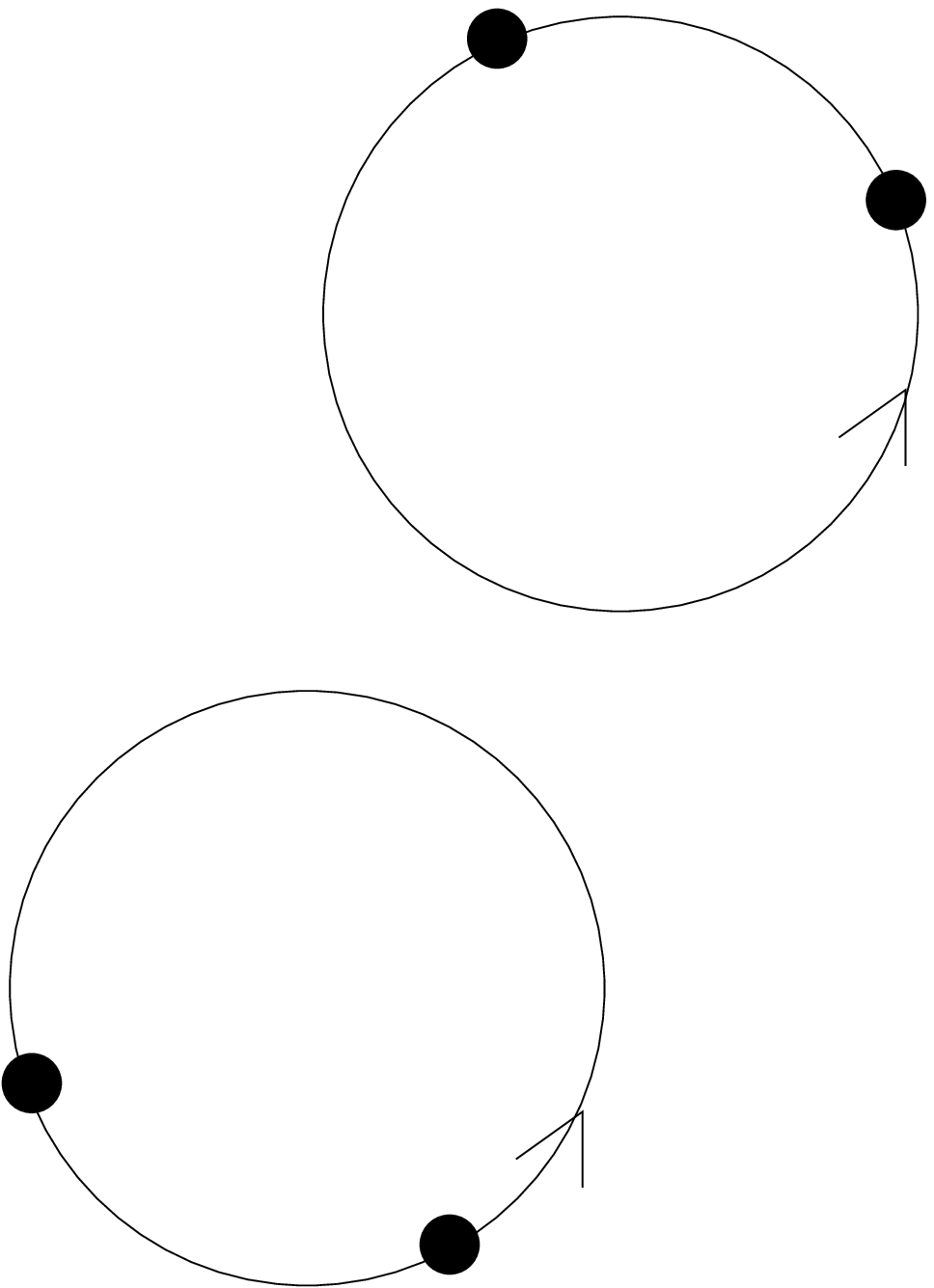}}\
=\, {2\over N}\,\vcenter{\epsfxsize=0.13\hsize\epsfbox{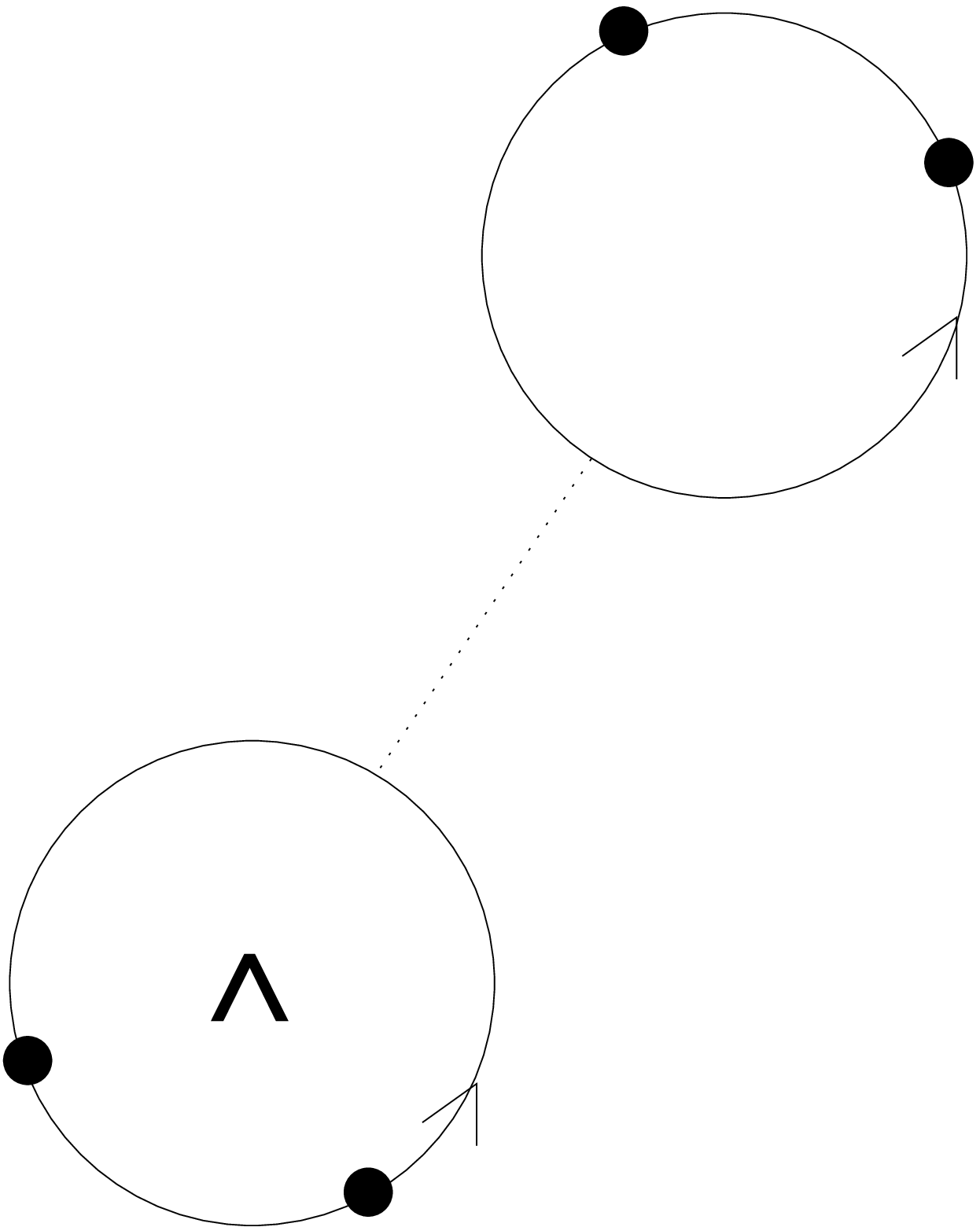}}
-{1\over N}\,\vcenter{\epsfxsize=0.13\hsize\epsfbox{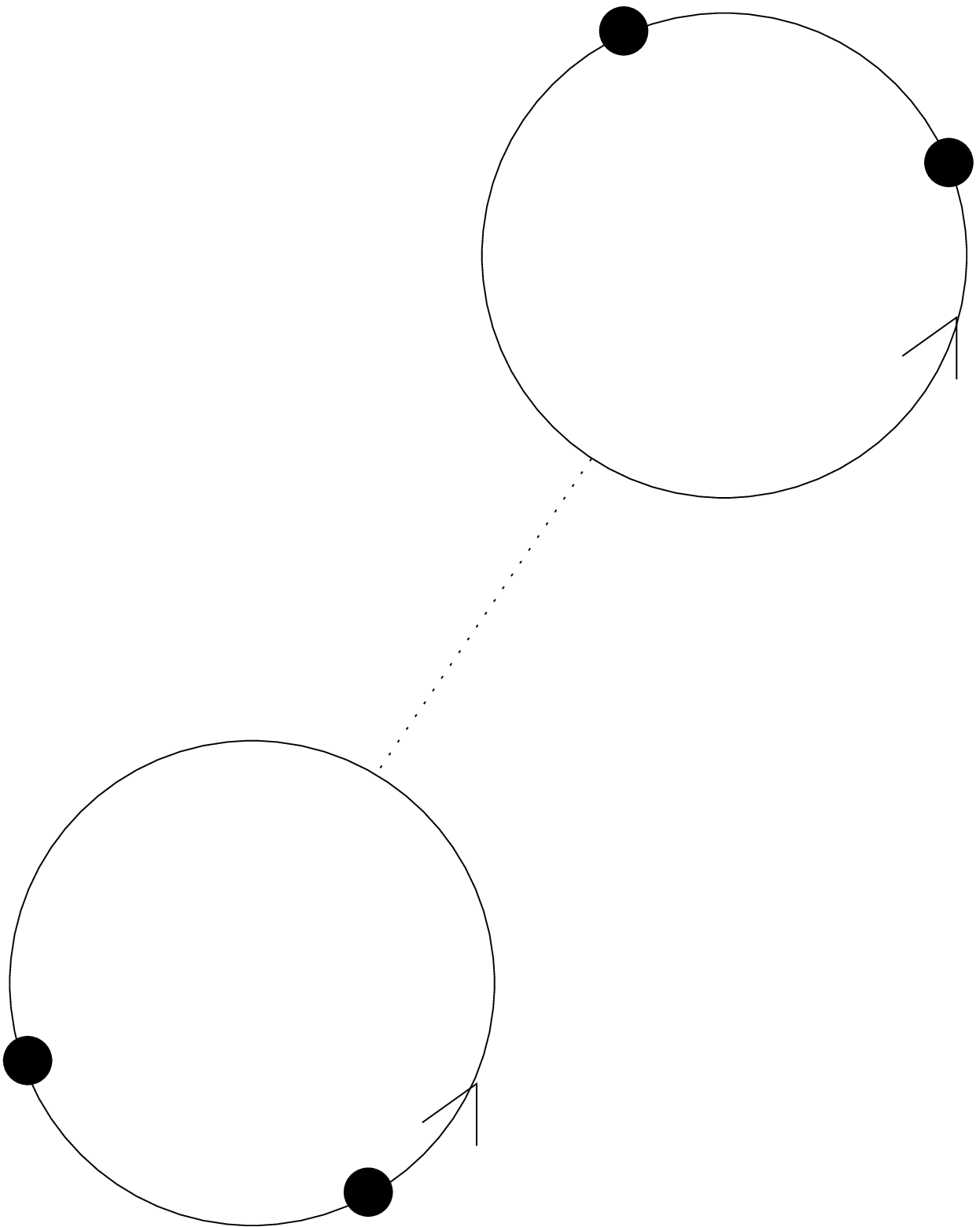}}
$$
\centerline{ 
{\bf Fig.13.} Feynman diagrams for the double-trace four-point vertex.}
\nfig\fdtr{Combine 390 and 210 for flow of $S^0_{\mu\nu,\la\si}(p,q;r,s)$
but do not forget $S_0$ and $\hS$ parts}
These contributions
vanish on adding the cyclic permutations $p^\mu\leftrightarrow q^\nu$ or
$r^\la\leftrightarrow s^\si$, by the antisymmetry of the three-point
vertices [as in \examsy]. Therefore $S^0_{\mu\nu,\la\si}(p,q;r,s)$ is
independent of $\Lambda$. Since by dimensions it is dimensionless, by
gauge invariance orthogonal to $p^\mu$, $q^\nu$, $r^\la$ and $s^\si$,
and by locality polynomial in these momenta, the only solution is 
$S^0_{\mu\nu,\la\si}(p,q;r,s)=0$. Clearly in this way one may
establish that 
$S^0_{\mu\nu,\la_1\cdots\la_n}(p,q;k_1,\cdots,k_n)$ vanishes
for any number $n$ of gauge fields inside the second trace.
(This remark is of importance for the large $N$ limit of the flow equation,
as already mentioned in sect. 5.) The first 
non-vanishing double-trace vertex appears at the six-point level,
with three gauge fields inside each trace.

Incidentally, we can now use \opcl\ to confirm the statement above
\deforg, that apart from ${\rm tr}\!\int\!\!d^D\!x\,
F_{\mu\nu}^2$  there are no relevant or marginal
operators here (as expected).  We see by expanding \opcl\ into $n$-point
vertices and using
\twop, that the term $\O_{min}$ in $\O_0[A]$, with the lowest number $m_{in}$
of gauge fields, cannot have any $\Lambda$ dependence. Thus 
the (eigen)dimension of $\O$ (which about the Gaussian fixed point $g=0$, is
that of $\O_0$) is simply $d_\O=m_{in}+m_p$,
where $m_p$ is the power of momentum in $\O_{min}$ (which must be a
positive integer by quasilocality).
Taking into account the requirement of gauge invariance we have 
that apart from the $m_{in}=m_p=0$ unit operator (\aka vacuum energy) 
which we here
always ignore, and the $m_{in}=m_p=2$ part of ${\rm tr}\!\int\!\!d^D\!x\,
F_{\mu\nu}^2$ which by \defg\ amounts to a change of $g$,
there are no operators with $d_\O\le4$.

\subsec{The $\beta$ function}

As usual, the $\beta$ function is determined through a renormalisation
condition, which in our case is \defg. Using this we have
\eqn\rencon{S_{\mu\nu}(p)={2
/ g^2}\,\Delta_{\mu\nu}(p)+O(p^3)\quad,}
and thus by \twop,
$$S_{\mu\nu}(p)={1\over g^2}S^0_{\mu\nu}(p)+O(p^3)\quad.$$
By \Sloope, this implies {\sl that the $O(p^2)$ component of
all the higher loop contributions $S^n_{\mu\nu}(p)$,  must vanish}.
This greatly simplifies the $O(p^2)$ part of the two-point vertex flow in
\ergone\ -- \ergtwo,  in particular reducing them to algebraic equations.
Thus we see that
\eqn\betaone{ 
 a_1[S_0-2\hS]_{\mu\nu}(p) =-4\beta_1\Delta_{\mu\nu}(p)+O(p^3)\quad,}
where $a_1[S_0-2\hS]_{\mu\nu}(p)$ is the two-point
vertex in $a_1[S_0-2\hS]$. This fixes $\beta_1$. Similarly,
at $n\ge2$ loops,  the $\beta_n$ are determined by the requirement that  
$a_1[S_{n-1}]_{\mu\nu}(p)=-4\beta_n\Delta_{\mu\nu}(p)+O(p^3)$.
And non-perturbatively from \ERG\ and \Sloope,
$$a_1[g^2S-2\hS]_{\mu\nu}(p)=
-{4\over g^3}\beta(g)\Delta_{\mu\nu}(p)+O(p^3)\quad.$$

\subsec{One loop}

We can use the above classical vertices to compute the one-loop 
two-point vertex and thus  $\beta_1$.
The relevant diagrams for the LHS of \betaone\foot{corresponding to $\Lambda\partial 
S^1_{\mu\nu}(p)/\partial\Lambda-2\beta_1 \hS_{\mu\nu}(p)$,
by \twop\ and \ergone.} are displayed in \fig\fbeta{fig.14
Alg -- the one loop beta1 diags except now Sigma stands for $S_0-2\hS$,
and extra $1/N$'s appear, also indicate the k direction}, 
where the circle stands for $\Sigma_0=S_0-2\hS$.
$$
{2\over N}\ \vcenter{\epsfxsize=0.085\hsize\epsfbox{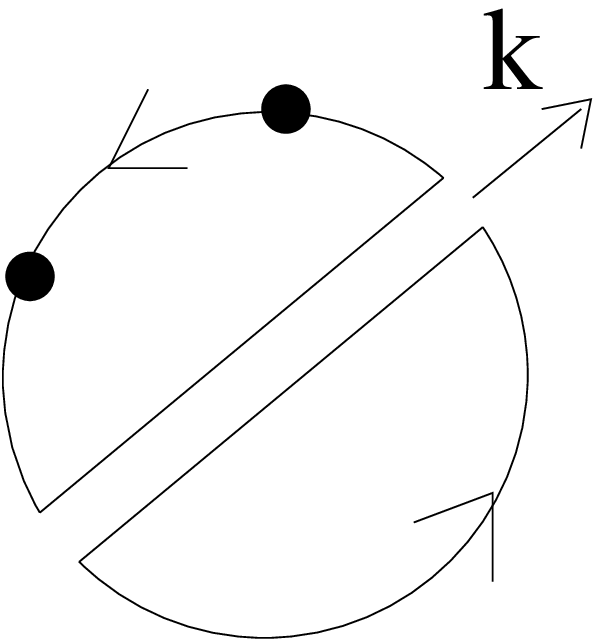}}
+{2\over N}\ \vcenter{\epsfxsize=0.085\hsize\epsfbox{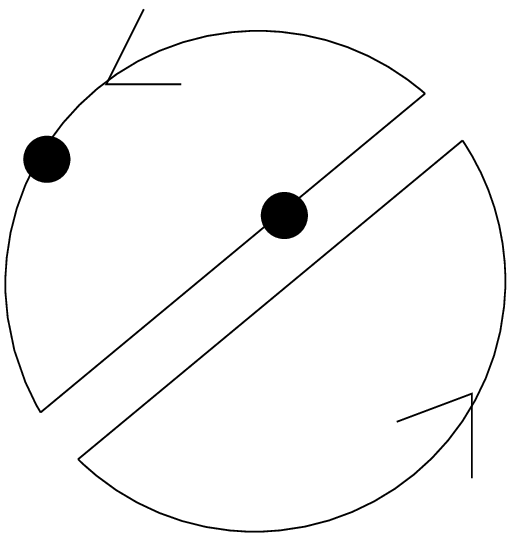}}
+{2\over N}\ \vcenter{\epsfxsize=0.085\hsize\epsfbox{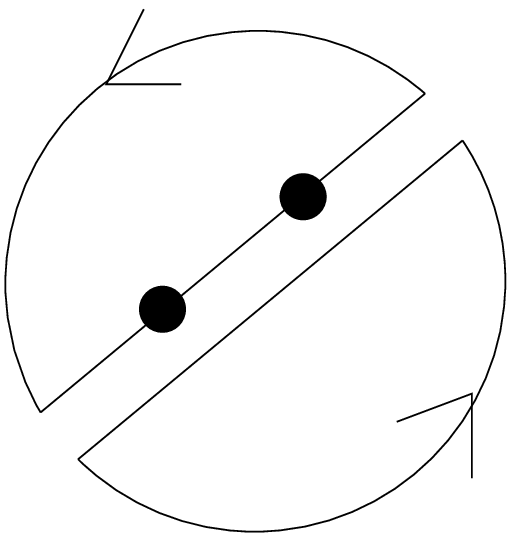}}
-{2\over N^2}\ \vcenter{\epsfxsize=0.085\hsize\epsfbox{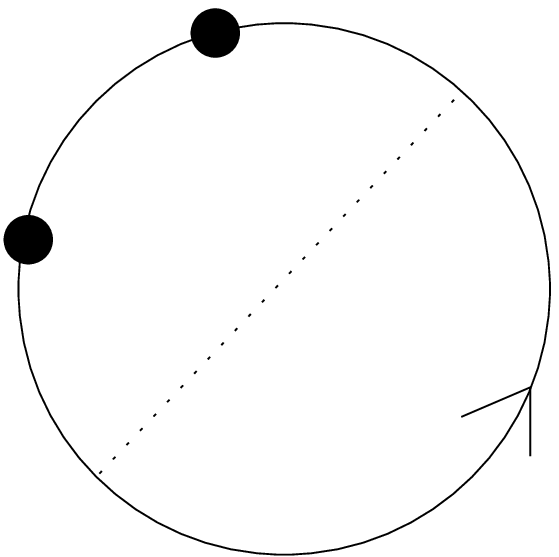}}
-{1\over N^2}\ \vcenter{\epsfxsize=0.085\hsize\epsfbox{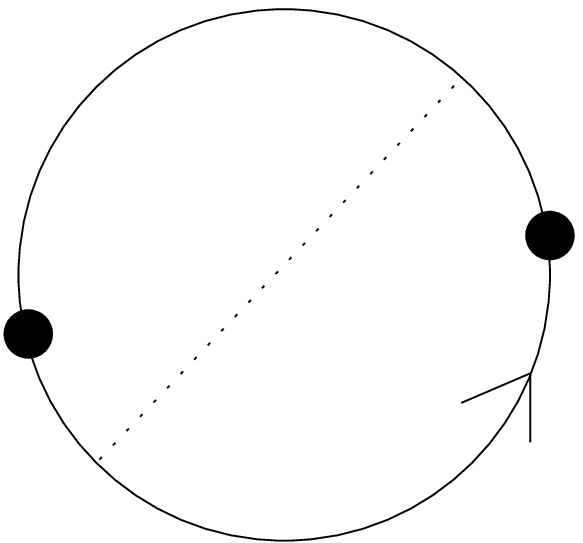}}
$$
\centerline{ 
{\bf Fig.14.} The one-loop two-point diagrams.}
Translating the Feynman rules, we get \eqnn\ol
$$\displaylines{a_1[\Sigma_0]_{\mu\nu}(p) =
{1\over\Lambda^2}\int\!{d^Dk\over(2\pi)^D}\Bigg\{
c'_k\Sigma^0_{\alpha\mu\nu\alpha}(\k,\p,-\p,-\k)
+c'_\mu(\p;\k-\p,-\k)\Sigma^0_{\alpha\nu\alpha}(\k,-\p,p\!-\!k)\hfill\cr
\hfill+\Sigma^0_{\alpha\alpha}(k)c'_{\mu\nu}(\p,-\p;\k,-\k)
-{c'_k\over N^2}\left[\Sigma^0_{\alpha\mu\nu\alpha}(k,p,-p,-k)+{1\over2}
\Sigma^0_{\alpha\mu\alpha\nu}(k,p,-k,-p) \right]
\Bigg\}\hfill\cr
\hfill +\left(p^\mu\leftrightarrow-p^\nu\right)\quad.\qquad\ol\cr}$$
Clearly by the Lorentz symmetry of the integral, the addition of the
cyclic permutation $p^\mu\leftrightarrow-p^\nu$ just multiplies the result
by two. Untying the large-$N$ change of variables $g\mapsto g/\sqrt{N}$,
we see that $\beta_1$ in principle has $N$ dependence in two
terms, one proportional to $N$, as expected,
and one unexpected contribution proportional to $1/N$.\foot{The coefficient
of each power of $N$ in \ol\ is separately completely constrained by gauge 
invariance, up to an overall factor, in the sense outlined below \fourp.}
Actually this $1/N$ term vanishes independently of the choice
of covariantization.  This follows since by using \hSex,
Lorentz invariance ($k\leftrightarrow-k$, 
$\mu\leftrightarrow\nu$ \etc) and the
last (coincident line) identity in \examsy,
one may show that under the $k$-integral
$\hS_{\alpha\mu\alpha\nu}(k,p,-k,-p)\equiv-2\hS_{\alpha\mu\nu\alpha}
(k,p,-p,-k)$,
while expanding the $S_0$ four-point vertex by \fourp, and again using 
these symmetries, one may show that the same identity holds for 
$S^0_{\alpha\mu\alpha\nu}(k,p,-k,-p)$.

The problem is that the $k$ integral 
is divergent. While propagator-like terms have been UV improved to
$\sim c/p^2$, the interactions \eg \hSex, \threep\ and \fourp, are worse
by $c^{-1}$, this latter behaviour being forced by gauge invariance \ga,
as we show in appendix A
\alg. With a power-law regulator,
\ie $c(x)\sim1/x^r$, $r>0$, for large $x\equiv k^2/\Lambda^2$, the integrand
$\sim1/k^2$ for large $k$. This behaviour may be established by
analysing the individual contributions, using the large $k$ behaviours
given in appendix A. It is most readily 
seen to hold in the $a_1[\hS]_{\mu\nu}(p)$ contributions:
The $n$-point
$\hS$ vertices with two large momenta $\sim\pm k$  behave as $c^{-1}_k
k^{4-n}$, while the $m$-point $c'$ vertices 
with $\sim\pm k$ entering in at their `ends' [as in \ol]
behave as $c'_k k^{-m}$. For the two-point one-loop vertex, we have
$m+n=4$.

On taking into account
the two powers of external momenta that must fall out by gauge
invariance, as in \betaone, one might expect that
the integrands behaviour is improved to $\sim1/k^4$, however
this is false. 
Firstly, this power counting argument relies on dimensional reasoning
which fails to work so simply here because it does not constrain
functions of the dimensionless ratios $\sim$momentum$/\Lambda$.
Secondly there is no separation of tree-level and loop 
contributions. These points are illustrated below.
Finally, gauge non-invariant terms may also arise,
from contributions that
cancel, but only after a shift in $k$ \alg.  These contributions
thus integrate to
surface terms in \ol, which are guaranteed to 
vanish only if the integral over the original parts converge.

To demonstrate this effect we compute the gauge dependent part of \ol.
The momentum routing chosen in \fbeta\ and \ol\
corresponds to taking $k$ to be the momentum of the left-most functional
derivative in the quantum term of \RG\ and
is natural if we wish to further regularise the integral in a way that
corresponds to modifying the measure over $A_\mu(k)$ modes whilst
preserving the fact that \RG\ leaves the partition function \zeff\
invariant, and thus preserve the expected universality of $\beta_1$.
However the reader can check that $k$-shifted differences always arise,
but as differing expressions, 
no matter how the terms in \ol\ are routed. (This reflects
the conflict between specifying momenta, and the action of 
gauge invariance \qap\alg.)
We assume that \ol\ is
further regularised so that invariance under $k
\leftrightarrow-k$ is preserved, and collect terms using the 
observation above \coids.
Then using \ga, \wga, and \twop, we readily obtain 
\eqn\gilch{p^\mu p^\nu a_1[\Sigma_0]_{\mu\nu}(p)=
{2\over\Lambda^2}\int\!{d^Dk\over(2\pi)^D}\left\{ c'_{k}
\hS_{\alpha\alpha}(k)-c'_{k+p}\hS_{\alpha\alpha}(k\!+\!p)\right\}\quad.}
Using \hSex\ and expanding the RHS as a power series in $p$,
one obtains integrals that may be done exactly and thus, using \betaone,
\eqn\gad{\eqalign{a_1[\Sigma_0]_{\mu\nu}(p) &=-4(D-1)\Omega_D\Lambda^{D-4}
\left\{\Lambda^2\delta_{\mu\nu}\left[G_0\right]^\infty_0
+p_\mu p_\nu\left[G_L\right]^\infty_0
\right\} -4\beta_1\Delta_{\mu\nu}(p)+O(p^3)\cr
{\rm where}\qquad G_0 &={1\over D}x^{D/2}\left({xc'/ c}\right)'
\qquad{\rm and}\qquad G_L={1\over D(D+2)}\left[x^{D/2+1}\left(xc'/c\right)''
\right]'\quad.
}
}
Here $\Omega_D=2/[\Gamma(D/2)(4\pi)^{D/2}]$ is the solid angle of a 
$(D-1)$-sphere divided by $(2\pi)^D$, and as before prime is 
differentiation with respect to its argument (here $x$).
These expressions give some guide as to the typical divergences expected
using the cutoff function $c(x)$. In particular for the power-law 
$c$, we have 
\eqn\lr{x c'/c\to -r\ins11{as}x\to\infty\quad,}
and the leading divergences actually cancel. If in addition the correction
to \lr\ is $O(1/x)$ or better (as will be the case here), 
the non-invariant terms in \gad\ vanish for all $D<4$, and thus also
for $D\to4^-$.  Note that {\sl at} $D=4$, we could still
get a finite $O(p^0)$ term from the subleading correction to \lr,
violating gauge invariance.
This illustrates the need to define these integrals carefully.
As we will see below however, power-law $c$ is not enough to
regularise the gauge-invariant divergences.

Consider \fig\fperpi{Use example beta of p644. Explain curved wine
comes last ... \ie the $\Lambda$ integrations}. This is
one of several $\sim1/k^2$
contributions to the first term of \ol\
(from expanding the first term in \ffour, using \fthree)
which, since the two blobs appear on two-point $\hS$ vertices,
are already orthogonal to $p$. Another example is given in 
ref.\alg. 
\midinsert
$$
\epsfxsize=0.4\hsize\epsfbox{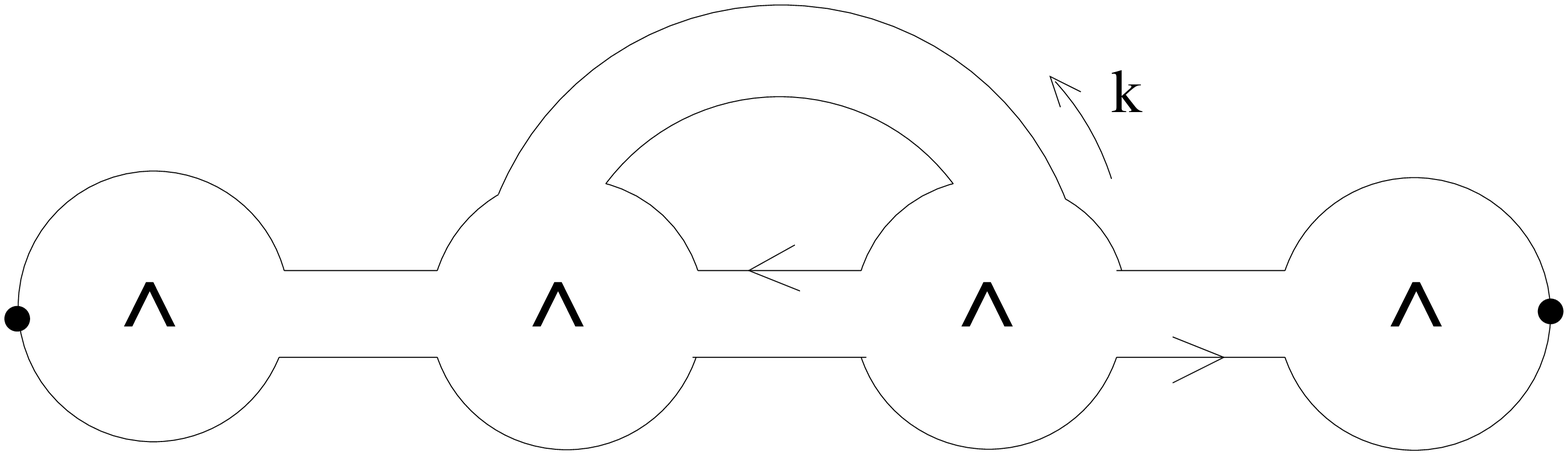}
$$
\centerline{{\bf Fig.15.} Divergent one-loop contribution
(made one-loop by the curved wine).}
\endinsert
These outer two-point vertices do not feel the loop-momentum $k$,
nor indeed do the outer wines,
and in this sense are really tree corrections to the one-loop
term. Indeed, however many points appear on these outer parts
the divergence is still quadratic in $D=4$ (\ie 
integrand goes as $\sim1/k^2$). 
Note that the two outer wines  
external momentum dependence 
$c'_p$ ensure that the coefficient of this
divergence is even non-polynomial of the form
$$\int_\Lambda^\infty\!\!{d\Lambda_1\over
\Lambda_1^3}\cdots\left(\int_{\Lambda_1}^\infty\!\!{d\Lambda_2\over
\Lambda_2^3} c'_p c^{-1}_p\cdots\right)^2\quad.$$
(The non-polynomial part arises from the $c'_p$ and it
should be understood that the inner $c$'s are functions of $\Lambda_2$.)
It is easy to verify either diagrammatically or directly from \threep,
\fourp\ and \ol, that
no other contribution has the same non-polynomial
dependence, and thus this divergence remains uncancelled.
Using appendix A, we isolated exactly the complete coefficients 
(polynomial and non-polynomial) of the
leading (and sub-leading) divergences, and verified that there
are many further uncancelled divergences. In the interest of
compactness we do not present the details.

\newsec{Conclusions}

We briefly recapitulate some of the main points. 

We formulate a  flow equation for the effective action $S$ of a
non-Abelian gauge field $A_\mu$,
the connection for the covariant derivative $D_\mu=\partial_\mu-iA_\mu$, 
which leaves the partition function invariant under the flow.

Gauge invariance is manifest at all stages; the
transverse bilinear term in $S$ need not be inverted
in the equation or its 
solution (which is here investigated perturbatively). Ghosts and gauge fixing
are not required. The challenging problem of Gribov copies
is thus avoided. 

({\it N.B.} Contrary to popular folklore, it is not necessary to
implement gauge fixing or other explicit reduction of variables merely
to factor out the infinite volume of the gauge group, which in any case
only amounts to one of many harmless divergent vacuum energy contributions.) 

The placing of the coupling
constant $g$ outside the connection, together with the exact preservation
of gauge invariance, means that $A_\mu$ is protected from wavefunction
renormalization. Only $g$ renormalizes. We show how the corresponding
$\beta$ function is determined through the appropriate renormalization
condition on $S$.

The global group invariance ensures that everything can be expanded in
products of traces. We show that in the large $N$ limit, $S$ collapses
to a single trace. The limit of the flow equation itself may be more
subtle. 

The general covariantization entering into the equation can be conveniently
expressed as an integral over configurations of Wilson lines. In a similar
way the solution $S$ may be expressed as an integral over a `gas' of Wilson
loops. In this way the flow equation may be reformulated entirely in terms
of manipulations on Wilson loops. The equations then determine
the measure over these fluctuating Wilson loops. In the large $N$ limit,
$S$ becomes an integral over the configurations of just one Wilson loop, \ie
 a path integral for a single
particle circulating in a loop. 

This Wilson loop picture may also be interpreted as Feynman diagrams and
thus provides an intuitive and elegant way to derive the perturbative 
solution for $S$. We derive the `Ward identities' and discuss the 
interpretation
of the various symmetries 
(charge conjugation, Lorentz and gauge invariance) in this picture.
We derive the two, three and four point vertices at the classical
level, and the one-loop two-point vertex.

We explicitly confirm that these solutions are gauge invariant. However,
we show that subtleties with momentum-integral surface terms appear at
one loop (and higher) as a consequence of the divergent nature the
momentum integrals in this framework. This in turn is related to
the large momentum behaviour of the vertices, whose lower bound 
is fixed by the exact preservation of
gauge invariance (as we show in appendix A). We demonstrate that 
some of the
divergences that thus appear at one loop are non-polynomial in momenta
and remain uncancelled in the complete solution in this framework.
The further regularisation that is required is developed in ref.\ymii.

The flow equation and its solution $S$ enjoy the important property
of `quasilocality', namely that they can be expanded in a Taylor 
series in their (external) momentum arguments to all orders. Equivalently
for $S$, that it has a derivative expansion to all orders. The 
completely regularised flow equation will correspond to integrating
out high energy modes because $\Lambda$ will then act properly as an
effective ultraviolet cutoff. The preservation of the partition function
then ensures that those high energy modes that are eliminated, get 
incorporated into $S$. 
The solution to the flow equation thus yields for the first time, a 
precise
continuum prescription for
a gauge invariant Wilsonian effective action.

\bigbreak\bigskip\bigskip\bigskip\bigskip
\centerline{{\bf Acknowledgements}}\nobreak
The author acknowledges support of the
PPARC through an Advanced Fellowship, and grant GR/K55738.

\bigbreak\bigskip\bigskip

\appendix{A}{Large and small momenta}
Any vertex has
a transverse part whose form
depends on the choice of covariantization $\{~\}$
(and the choice of flow equation in the case of $S$),
and a universal longitudinal part whose form is dictated by gauge
invariance. Consider the one-point wine $W_\mu(p;k,-k-p)$. (Recall that
we label
the one-point vertex by the kernel it covariantizes.
Since, by translation invariance, 
all vertices are accompanied by momentum conserving delta
functions, all relations must be derived for the case where the momentum
arguments sum to zero.) By gauge invariance \ga, \wga,  we have \examsy:
\eqn\gaw{p^\mu W_\mu(p; k, -k-p)=W_{k}-W_{k+p}\quad.}
In the case that $p$ and $k$ are colinear, this may be solved immediately
for $W_\mu(p;k,-k-p)$. 

More generally, for small $p$, using $W_k\equiv W(k^2/\Lambda^2)$, 
we have
\eqn\pw{W_\mu(p;k,-k-p)=-{W'_k\over\Lambda^2}(2k_\mu+p_\mu)
-2{W''_k\over\Lambda^4} k.p\, k_\mu +O(p^2)\quad,}
where we have noted that any term perpendicular to $p^\mu$ may be
added while leaving \gaw\ unchanged, but by Lorentz invariance
this has tensor structure $\Delta_{\mu\nu}(p)k^\nu$.

\gaw\ also constrains 
the ultraviolet behaviour of $W_\mu(p;k,-k-p)$.
For example, for large $k$ it 
is constrained to be at least as bad as the RHS.
Explicitly, consider choices of $W_k$
where the scale is set by $k$ for large momenta, and thus
$W'(x)/W(x)\sim1/x$ for large $x$,
\ie where 
$W_k$ goes as a power or log for large
$k$ (say $W(x)\sim x^w$). 
\pw\ then typically holds also for $k>\!\!>p$, with $O(p^2)$ 
replaced by $O(p^2 k^{2w-3})$, since for this to be violated
requires including non-minimal field strength 
terms\foot{balanced by $1/\Lambda^2$:
these are the terms which are transverse to $p$}
in $\{~\}$.

Many similar relations can be derived for colinear, small and large
momenta for the other vertices, including $\hS$ and $S$ vertices.
In any case we have that the ultraviolet behaviour
for large $k$ is at best $W_\mu(p;k,-k-p)\sim W_k/k$.
It is then easy to see that $W_{\mu_1\cdots\mu_n,
\nu_1\cdots\nu_m}(p_1,\cdots,p_n;q_1,\cdots,q_m;k,-k-\sum_i p_i-
\sum_jq_j)$ {\it must have ultraviolet behaviour at least as bad as }
$W_k/k^{m+n}$: By the gauge relations \ga, \wga, each $m+n$
vertex is related to a difference of two $(m+n-1)$ vertices,
whose arguments only differ in 
momenta small with respect to $k$.
For choices $W_k$ where $k$ sets the scale (as above),
this difference 
falls by just one extra power of $k$ compared to each $(m+n-1)$ vertex,
and thus the $m+n$ vertex cannot fall faster than an ($m+n-1$ vertex)$/k$.
The italicized statement above then follows immediately, by iteration.

Recall that the two-point vertices for the actions are fixed as in \twop\
and \hSex, and therefore $S_{\mu\nu}(k)=\hS_{\mu\nu}(k)\sim c^{-1}_k k^2$.
By the same argument as above, using \ga,
we have that the $n$-point
vertices $\hS_{\mu_1\cdots\mu_n}(p_1,\cdots,p_n)$ where any two
momenta $p_i$ are large of order $k$, behave at best as
$c^{-1}_k k^{4-n}$. The same bound holds for $S$. These relations
are sufficient for the demonstration that the one-loop contribution is not
regularized (in sect. 8), and show that the lack of complete regularization
follows from the gauge invariance.

In fact, while the
$\hS$ vertices achieve the lower bound $c^{-1}_k k^{4-n}$
(as sketched below), $S$ does not,
as is already clear from the flow of the classical four-point function 
shown in \ffour.
We see there that if the two large momenta 
are next to each other, they can enter and leave a three
point lobe, thus
actually  $S_{\mu\nu\lambda\sigma}(k,-k-p-q,p,q)\sim c^{-1}_k k$.

This effect would not appear if the gauge invariant flow was for
one-particle irreducible vertices (a Legendre formulation) 
but the lack of separation
between tree-level and loop contributions 
(see also the end of sect. 8) seems to be a necessary
consequence of the exact preservation of gauge invariance. To see
this, recall that
all the terms in \ffour\ have to appear as shown,
by gauge invariance, given the first two terms. But attaching a 
wine to make a one-loop contribution out of this 
gauge invariant four-point vertex
produces both diagrams that are one-particle reducible and one-particle
irreducible.

The cases where one or both large momenta enter the side of a wine are
also more involved. We rewrite \gaw\ as
\eqn\gawk{k^\mu W_\mu(k; p, -k-p)=W_{p}-W_{k+p}\quad.}
Now the asymptotic behaviour for large $k$ depends on whether $W_k$
grows or falls for large $k$. 
The growing case corresponds to $c^{-1}_k\sim k^{2r}$ 
and does still follow the
simple rules above, if $r$ is large enough: 
We have that at best $c^{-1}_\mu(k; p, -k-p)\sim c^{-1}_k/k$.
Arguments as above then rapidly establish that an $m+n$ point vertex
from $\{c^{-1}\}$ with two large momenta $\sim k$ 
in {\sl any} position behaves 
at best as $\sim c^{-1}_k/k^{m+n}$ providing $r\ge m+n$. 
And once again, this bound is saturated if there are no
non-minimal terms in $\{~\}$.
In the cases where $r<m+n$, the bound established below for 
the falling cases may dominate.

The falling case corresponds to $W_k=c'_k$ and in this case we
read from \gawk, that at best 
$c'_\mu(k;p,-k-p)\sim c'_p/k$. Note that the coefficient of a 
divergence arising from this need not be polynomial in $p$.
(See similar remarks in sect. 8.) This $1/k$ behaviour is true
of any $m+n$ vertex where the $k^\mu$ can reach the top end of the
line, for example:
$$k^\mu c'_{\lambda\mu,\nu}(r,k;q;p,-k-p-q-r)=
c'_{\lambda,\nu}(r;q;p,-p-q-r)-c'_{\lambda,\nu}(r+k;q;p,-k-p-q-r)\quad.$$
The first term on the RHS $\sim k^0$ sets the bound, since the second
term can go as $\sim1/k$ [as may be established by writing out the
gauge relation $k^\lambda c'_{\lambda,\nu}(k;q;p,-k-p-q)$]. 
Again by iteration, the general rule for the case
where one large momentum $k$ enters via the side of $\{ c'\}$ is that
the vertex goes at best as $1/k^{v+1}$ where $v$ is the number points 
(gauge fields) separating the two large momenta. Once more, this 
is saturated by minimal covariantizations. 

The case where both large momenta enter $\{ c'\}$ via the side can be
treated similarly, with the conclusion that the vertex again goes at
best as $1/k^{v+1}$ where $v$ is the number points 
separating the two large momenta. However in this case even minimal
covariantizations do not necessarily reach this bound. For \ourchoice,
the case where the two momenta are next to each other on one side of
the wine leads to a $O(k^0)$ contribution\foot{Note that it is necessarily
transverse, otherwise the $k^\mu$ gauge relations would be violated.} 
$\sim \Delta_{\mu\nu}(k)/k^2$ as will be established explicitly in
ref. \ymii. Consequently \ourchoice\ actually behaves as $1/k^v$,
for two large momenta $\sim k$ entering the side and separated by
$v$ points. 

As indicated above if $r<m+n$, in particular if $m+n-r>v+1$
($m+n-r>v$), then the 
bounds $\sim 1/k^{v+1}$ ($\sim 1/k^v$) for one (two) large side 
momenta are the limiting ones for $\{ c^{-1} \}$ also.
With the above bounds for wine vertices,
one can check explicitly from \hSex, that the $\hS$
vertices saturate their own bounds.
All the asymptotic formulae given in this appendix
will be confirmed explicitly for \ourchoice\ in ref. \ymii, where
we will also derive the coefficient functions multipying the leading
$k$ behaviour.

\listrefs

\end